\documentclass[letterpaper,peerreview,draftclsnofoot,onecolumn,11pt]{IEEEtran}

\usepackage{amsmath, amssymb, graphicx, cite, xcolor}

\newtheorem{lem}{Lemma}
\newtheorem{prop}{Proposition}

\DeclareMathOperator{\erf}{erf}

\DeclareMathOperator{\argmax}{\arg\,\max}

\begin{document}
\title{Queuing Approaches to Principal-Agent Communication under Information Overload}

\author{Aseem~Sharma,
        Krishna~Jagannathan,~\IEEEmembership{Member,~IEEE},
        and~Lav~R.~Varshney,~\IEEEmembership{Senior Member,~IEEE}
\thanks{A.~Sharma and K.~Jagannathan are with the Department of Electrical Engineering, Indian Institute of Technology Madras, Chennai 600036, India (e-mail: \{ee12s001, krishnaj\}@ee.iitm.ac.in).}%
\thanks{L.~R. Varshney is with the Coordinated Science Laboratory and the Department of Electrical and Computer Engineering, University of Illinois at Urbana-Champaign, Urbana, IL 61801, USA (e-mail: varshney@illinois.edu).}
\thanks{The material in this paper was presented in part at the 2014 IEEE International Symposium on Information Theory \cite{SharmaJV2014}.}
\thanks{This work was supported in part by NSF Grant CCF-1623821 (CIF: EAGER: Towards an Information Theory of Attention).}
}

\maketitle

\begin{abstract}
In the information overload regime, human communication tasks such as responding to email are well-modeled as priority queues, where priority
is determined by a mix of intrinsic motivation and extrinsic motivation corresponding to the task's importance to the sender.
We view priority queuing from a principal-agent perspective, and characterize the effect of priority-misalignment and information asymmetry between task senders and task receivers
in both single-agent and multi-agent settings.  In the single-agent setting, we find that discipline can override misalignment.
Although variation in human interests leads to performance loss in the single-agent setting, the same variability is useful to the principal with optimal routing of tasks, if the principal has suitable information about agents' priorities. 
Our approach starts to quantitatively address the effect of human dynamics in routine communication tasks.
\end{abstract}

\begin{IEEEkeywords}
Information overload, principal-agent problem, queuing
\end{IEEEkeywords}

\section{Introduction}
\label{sec:intro}
Many emerging informational technologies such as social media and collaborative productivity platforms are built on top of near-optimal technical solutions to reliable information flow \cite{CostelloF2007}, but there is little understanding of information-theoretic limits or optimal designs of these engineering systems themselves.  This is because such systems are not purely technical but sociotechnical in scope, where humans cannot be abstracted away \cite{WeckRM2011}.  To optimally design such systems, 
we may need a perspective that not only merges information-theoretic and network-oriented approaches 
\cite{Gallager1985, EphremidesH1998}, but also takes certain aspects of human behavior 
into account \cite{GinoP2008}.

In the accompaniment to Shannon's seminal work, Warren Weaver talks about different levels 
of communication, culminating with the so-called \emph{effectiveness problem}
\cite{ShannonW1949}: how effectively does the received meaning affect conduct in the desired way?  
In sociotechnical information systems, solving the effectiveness problem in human response to received information is key. 

Indeed the effectiveness of communication is strongly governed by limited human attention \cite{Kahneman1973}, as well as intrinsic and extrinsic motivating factors \cite{GagneD2005}. Our present work takes nascent steps to develop a mathematical understanding of sociotechnical communication, considering a \emph{principal-agent} formulation of a \emph{human priority queue}.\footnote{One might wonder whether human behavior is consistent enough to justify analysis through (stochastic) mathematical models, the way physical communication channels and information sources seem to be. We believe this is justified, as many long-standing results from psychology are consistent and dependable, displaying test-retest reliability, inter-rater reliability, parallel-forms reliability, and internal consistency reliability \cite{KaplanS2009}.}


Since there is a limit to the rate at which humans can work \cite{Kahneman1973}, tasks such as responding to emails queue up, especially in the information overload regime \cite{Spira2011}.  Also, humans generally do not perform their tasks in the order in which they are received, but act on them based on certain priorities \cite{schwartz1978queues}. Indeed, studies have shown that human communication dynamics, such as email correspondence, follow heavy-tailed timing distributions, which is in contrast to traditional queuing models that assume Poisson statistics, and suggests that patterns of deliberate human activity are at work. A prominent explanation for this empirical phenomenon is a priority queue model of human action \cite{Barabasi2005}.

What factors determine priorities? The information gap theory of curiosity \cite{Loewenstein1994} has often been demonstrated: people read an email if the subject line suggests it is probably communicative in Shannon's sense of having uncertainty \cite{WainerDK2011,DabbishKFK2005}, cf.~\cite{Varshney2013}. As mentioned in \cite{DabbishKFK2005}, factors other than message importance (such as social considerations) also play a role in the way people respond to emails. More generally, both \emph{intrinsic} and \emph{extrinsic} motivations contribute to the eagerness and speed with which people engage in their tasks \cite{GagneD2005,OswaldPS2009,IsenR2005}. Intrinsic factors push people to act due to interest and satisfaction from the activity itself. Extrinsic factors, on the other hand, involve people drawing motivation from the extrinsic consequences of completing the task. 

Consider, for instance, how an employee of an organization typically acts upon assigned tasks. Although the arriving tasks have various levels of importance for the organization, the employee's level of interest in each task also plays a significant role in his prioritization. Therefore, there is a possible misalignment between the priorities of the organization and those of the employee. Further, there is usually an \emph{information asymmetry} between the task sender and the employee---the sender may not be aware of the intrinsic motivation that affects the priorities of the employee towards the assigned tasks.

\subsection{Queuing-theoretic principal-agent communication}

We look at the human priority queue as a principal-agent problem \cite{Ross1973}, concerning ourselves with the impact on the principal (task sender) of decisions made by the agent (task receiver). The principal-agent problem occurs due to misalignment of interests and information asymmetry between the parties in decision problems.

Translating this to the effectiveness problem, the transmitter is the principal who assigns jobs to the agent. The principal wants her agent to perform tasks in a particular priority order, which are driven by her interests. Under the information overload regime, the agent has many tasks to perform, and hence his tasks queue up. In addition, the agent has different levels of interest  in the queued tasks, which may be different from the principal's interests. Though aware of the principal's interests, the agent prioritizes the tasks according to some function of his intrinsic and extrinsic motivation factors. Further, the principal does not have precise knowledge of the agent's interests. This information asymmetry and conflict of interests gives rise to the queuing-theoretic principal-agent communication problem we study. 

In this paper, we use a priority queuing framework to study two related problems, namely (i) task execution by people, and (ii) task distribution in hierarchical human organizations involving several subordinate agents. We aim to quantify the impact of information asymmetry, of interest-misalignment, and of diversity of human motivations in such systems. 

Note that strategic communication has been studied in economic theory \cite{CrawfordS1982, DewatripontT2005, AghionT1997, BoltonD2013}, but without key engineering considerations such as queuing. These papers take a fairly simplified game-theoretic perspective, and when principal-agent issues arise, they are primarily concerned with contract design and the imposition of interpersonal authority within organizations.  On the other hand, we take a priority queuing approach, and are specifically interested in real-valued engineering performance criteria such as delay and throughput.  Further note that studies in mass communication are rarely quantitative \cite{Berlo1960}.

At the heart of our work lies the fact that limited human attention and factors of prioritization lead to novel mathematical problems not encountered previously in communication network design \cite{SrikantY2014}. Our study suggests that queuing theory, which plays a crucial role in the design and analysis of communication networks \cite{tassiulas1993dynamic}, is also a useful tool in characterizing sociotechnical information flow.    



%
To characterize the performance of the priority queue from the principal's perspective, we define a cost function as the average priority-weighted sojourn time of the queue. The sojourn time is the number of time steps from when a task enters the queue to the time it leaves. A large cost implies a large delay for high-priority tasks, which is undesirable for the principal.

We develop expressions for the cost function for various scenarios. We start with the analysis of a principal-agent problem with one agent. Under this setting, we first look at variation in agent's priorities due to extrinsic motivation and conflict of interests between principal and agent: we focus on understanding the effect of the alignment between the principal's and the agent's interests, as well as the effect of extrinsic motivation on the agent's prioritization.
Second, we compare an agent working at variable rate---the variation being a function of the agent's interest---with one working at a constant rate.
While we find that a variable agent performs worse than a constant agent in the single-agent case, we find a contrary result when extended to the multi-agent case under certain circumstances. In the multi-agent case, we explore the effect of information asymmetry and seek to find optimal routing policies that minimize the cost for all tasks. We extend that objective to find routing policies that minimize the cost for a given subset of tasks. 

The insights resulting from these models form important building blocks for designing information flow in sociotechnical systems. For concreteness and tractability, we use specific statistical assumptions (Gaussian distributed priorities, etc.), but the general modeling principles and insights should hold more broadly.  Even in these simplified settings, we get some non-intuitive and surprising mathematical results.

\section{System Model and Notation}
\label{sec:model}
We model agents as discrete-time preemptive resume priority queues. Time is slotted, with service initiation at slot boundaries. The scheduling discipline is \emph{preemptive resume}: a newly arriving task can interrupt the ongoing service of a task, and the interrupted task can resume service when the former has finished. In our case, preemption is based on task priority, i.e., a newly arriving task can interrupt the ongoing service of a task if the former has a higher priority than the latter. Arrivals in different time slots are i.i.d., with arrival and service processes independent of each other. To ensure stability, the average number of tasks arriving in the system is strictly less than the average service rate.

Let random variable $X$ denote the principal's interests, and let random variable $D$ denote the sojourn time of a task in the queue. The sojourn time of a task is the time from when the task enters the queue to when it leaves. The \emph{cost} measures performance and is the average priority-weighted sojourn time of the queue, $\mathbb{E}[XD]$. We initially assume a general distribution for arrival and service processes, but then look at particular distributions as examples.

For the single-agent case, let random variable $Y$ denote the agent's interests. Let $A(k)$ denote the number of tasks arriving in the queue in the $k$th time step
and let $A_z(k)$ denote the number of tasks arriving in the queue in the $k$th time step with priorities higher than $z$. Let $S$ denote the service time of a task. Let $\lambda$ be the arrival rate of the tasks in the queue. When the service rate is the same for all tasks, we call it $\mu$, where $\mu=1/\mathbb{E}[S]$.  When the service rate is variable, it is a function of the agent's interest in the task, $Y$. The cost for the single-agent case is denoted $C_o$.

For the multi-agent case with two agents, let random variables $Y_1$ and $Y_2$ denote their interests. Let $S_1$ and $S_2$ denote the service times of a task in the two queues. We consider \emph{memoryless} task allocation functions that depend on one or more parameters of the same task only, and not on parameters of other tasks. Let $\lambda_1$ and $\lambda_2$ be the arrival rates of the tasks in the queues, and let $\lambda=\lambda_1 + \lambda_2$ be the total arrival rate. When the service rate is the same for all tasks, we assume service rates $\mu_1$ and $\mu_2$, $\mu_i=1/\mathbb{E}[S_i]$. For variable service rate, the service rate of a task is governed by that agent's interest in the task. We denote the cost for the multi-agent case by $C_m$.

\section{Single-agent principal-agent communication}
Consider a single-agent principal-agent communication model, where the principal has one agent to allocate tasks to. Each task has two kinds of priorities, $X$ and $Y$. The $X$ term is the importance of a task to the principal whereas the $Y$ term captures how interesting the task is to the agent. The principal would want the agent to prioritize tasks according to $X$. The agent is aware of $X$, but since he has his own set of interests $Y$, he uses a function of both $X$ and $Y$ to prioritize the tasks. The correlation between $X$ and $Y$ indicates how aligned the principal and agent are.
We analyze two variations of the model:
\begin{itemize}
\item Priority variation with agent's interests: The agent serves each task at the same rate, but prioritizes the tasks as a function of the principal's and his own interests. Here, we are interested in analyzing the effect of interest misalignment and intrinsic motivation on the cost function. For the sake of concreteness and tractability, we model $X$ and $Y$ as jointly Gaussian random variables.
\item Service rate variation with agent's interests: The agent prioritizes the tasks according to the principal's interests, but serves each task at a different rate based on his own interest in the task. Here, we are interested in analyzing the effect of the agent's service rate on the cost function. In this case, we model $X$ and $Y$ as uniformly distributed.
\end{itemize}

\subsection{Priority variation}

Each task is served at the same rate. For concreteness, we consider $X$ and $Y$ as jointly Gaussian random variables with correlation coefficient $\rho$ (which completely captures the alignment between principal and agent). A task can have any real number as its priority: the lower the priority, the more negative its value. The agent prioritizes according to random variable $Z$, which is a linear function of $X$ and $Y$:
\begin{equation*}
Z=g(X,Y)\mbox{.}
\end{equation*}
A task with larger $Z$ is served first.

Recall the cost $C_o$ is the average priority-weighted sojourn time $C_o=\mathbb{E}[XD]$, where $D$ is the sojourn time of a task in the queue. Note that the cost takes $X$ but not $Y$ into account, reflecting the fact the principal's cost is governed only by $X$. Also note that the principal would want the cost of the agent queue to be as small as possible. A well-aligned agent will have negative cost for its queue, since lower priority tasks (which have negative priorities) will have larger sojourn times. Using the law of iterated expectations gives:
\begin{equation}
\label{eq:conditional cost}
\mathbb{E}[XD]=\mathbb{E}\left[\mathbb{E}[XD | Z]\right] \mbox{.}
\end{equation}
The sojourn time of a task is a function of $Z$, and thus depends on $X$ through $Z$. However,
$D$ is conditionally independent of $X$ (conditioned on $Z$). This follows from applying the law of
iterated expectations to \eqref{eq:conditional cost}:
\begin{align}
\label{eq:iterating cost}
\mathbb{E}\left[\mathbb{E}[XD | Z]\right] &= \mathbb{E} \left[ \mathbb{E}\left[\mathbb{E}[XD | Z,X ] | Z \right] \right] \notag \\
&= \mathbb{E} \left[ \mathbb{E}\left[ X \mathbb{E}[D | Z,X] | Z \right] \right]\mbox{.}
\end{align}
The sojourn time of a task $D$ is a random variable that is a function of the priority of that task in the queue. This implies $D$ is a function of $X$ and $Y$ only through $Z$. In other words, knowing $Z$, the distribution of $D$ does not change with knowledge of $X$:
\[
\mathbb{P}(D \leq d | Z) = \mathbb{P}(D \leq d | Z, X) \mbox{.}
\]
Therefore,
\[
\mathbb{E}[D | Z] = \mathbb{E}[D | Z, X] \mbox{.}
\]
Consequently, using \eqref{eq:iterating cost}:
\begin{equation*}
\mathbb{E} \left[ \mathbb{E}\left[ X \mathbb{E}[D \mid Z,X ] \mid Z \right] \right]=\mathbb{E} \left[ \mathbb{E}\left[ X \mathbb{E}[D \mid Z] \mid Z \right] \right]\mbox{.}
\end{equation*}
As a result,
\begin{equation}
\label{eq:iterated cost}
\mathbb{E}[XD]=\mathbb{E}\left[\mathbb{E}\left[X \mid Z\right]\mathbb{E}\left[D \mid Z\right]\right]\mbox{.}
\end{equation}

To find $\mathbb{E}[X|Z]$, we note that since $X$ and $Y$ are jointly Gaussian, $X$ and $Z$ will
also be jointly Gaussian. The conditional expectation of $X$ can thus be obtained as
\begin{equation}
\label{eq:conditional expected x}
\mathbb{E}[X|Z]={\mu}_{x} + {\rho}_{x,z}\frac{{\sigma}_{x}}{{\sigma}_{z}}(Z-{\mu}_{z})\mbox{,}
\end{equation}
where $\mu_x$ and $\mu_z$ are the respective means, $\sigma_x^2$ and $\sigma_z^2$ are the respective variances, and $\rho_{x,z}$ is the correlation coefficient.
The value of $\mathbb{E}[X|Z]$ can be explicitly obtained from the statistics of $X$ and $Y$,
and the function $g$.

Next we derive an expression for $\mathbb{E}[D|Z]$. Let us tag a task with $Z=z$. Since the
scheduling discipline is preemptive resume priority, this task's sojourn time is affected only by the
tasks whose priorities are greater than $z$, and not the order in which they are executed. Hence, we
can map our continuous priority queue onto a two-class priority queue. The tagged task constitutes
the low-priority class and all the higher priority tasks constitute the high-priority class. We can
now employ standard results from the priority queueing literature. From \cite{WalraevensDMB2012},
the expected sojourn time of a task with priority $z$ is
\begin{align}
\label{eq:conditional expected sojourn time}
\mathbb{E}[D | Z=z] = \frac{\left( 2\mu -{\lambda}_{z}\right) \mbox{var}\left( A_z(k) \right)}{2{\lambda}_{z}{\left( \mu - {\lambda}_{z} \right)}^{2}} 
+ \frac{{\lambda}_{z} {\mu}^{2} \mbox{var} \left( S \right)}{2{\left( \mu - {\lambda}_{z} \right)}^{2}} - \frac{{\lambda}_{z}}{2 \left( \mu - {\lambda}_{z} \right)}\mbox{,}
\end{align}
where $\lambda_z$ is the average number of task arrivals in a time step with priorities higher than $z$,
and var$(A_z(k))$ and var$(S)$ are the variances of the number of task arrivals in a time step with
priorities higher than $z$ and of the execution times, respectively.

Let $p_z$ be the probability that an arriving task has a priority greater than $z$. To find $\mathbb{E}[ D | Z=z ]$, we see that the variance values var$( A_z(k) )$ and var$(S)$ can be obtained from the statistics of the arrival and the service processes. Moreover, since $\lambda_z$ is the fraction of the average number of tasks that have priorities greater than $z$, we can write $\lambda_z=p_z \lambda$.
Hence, using \eqref{eq:conditional expected x} and \eqref{eq:conditional expected sojourn time} in \eqref{eq:iterated cost}, we obtain the following expression for the cost function
\begin{align}
\label{eq:complete cost}
&C_o=\int_z f_Z \left(z \right) \, dz \left( {\mu}_{x} + {\rho}_{x,z}\frac{{\sigma}_{x}}{{\sigma}_{z}}\left(z-{\mu}_{z}\right) \right)\\
&\times \left( \frac{\left( 2\mu -{\lambda}_{z}\right) \mbox{var} \left( A_z(k) \right)}{2{\lambda}_{z}{\left( \mu - {\lambda}_{z} \right)}^{2}} + \frac{{\lambda}_{z} {\mu}^{2} \mbox{var} \left( S \right)}{2{\left( \mu - {\lambda}_{z} \right)}^{2}} - \frac{{\lambda}_{z}}{2 \left( \mu - {\lambda}_{z} \right)} \right)  \mbox{.} \notag
\end{align}
To gain insight into this expression, let us evaluate it for a concrete example.
\subsubsection{Example}
Let $X$ and $Y$ be zero-mean, jointly Gaussian
random variables with correlation matrix
\begin{equation*}
K=\begin{bmatrix}
1 & \rho \\
\rho & 1 \\
\end{bmatrix} \mbox{.}
\end{equation*}
Here, $\rho$ denotes the extent of alignment between the principal and agent. When $\rho=1$, there is
perfect alignment between the principal's interest $X$ and the agent's interest $Y$. When $\rho=-1$, there is
perfect misalignment. When $\rho=0$, $X$ and $Y$ are independent.  We take the random variable $Z$ as
\begin{equation}
\label{eq:prioritization function}
Z=\frac{\gamma X + \left( 1-\gamma \right) Y}{\sqrt{{\gamma}^2 + {\left(1-\gamma \right)}^2 + 2 \rho \gamma \left( 1-\gamma \right)}}\mbox{,}
\end{equation}
where $\gamma \in [0,1]$ is a parameter that indicates how much importance the agent gives to the principal's interests. Whereas $\rho$ captures the intrinsic alignment of the agent with the principal, $\gamma$ captures the extrinsic motivation that aligns the agent with the principal. The denominator is the standard deviation of $\gamma X + \left( 1-\gamma \right) Y$. Thus, $Z$ is unconditionally a standard Gaussian random variable. Hence, \eqref{eq:conditional expected x} becomes
\begin{equation}
\label{eq:conditional expected x for mm1}
\mathbb{E}[X | Z]=\rho_{x,z} Z\mbox{,}
\end{equation}
where $\rho_{x,z}$ can be calculated as
\begin{align}
\label{eq:rho_xy}
\rho_{x,z}&=\frac{\mbox{cov}(X,Z)}{\sqrt{\mbox{var}(X) \mbox{var}(Z)}} \notag \\
&=\frac{\gamma + \rho \left( 1-\gamma \right)}{\sqrt{{\gamma}^2 + {\left(1-\gamma \right)}^2 + 2 \rho \gamma \left( 1-\gamma \right)}} \mbox{.}
\end{align}

For the specific case where the number of arrivals in a given time has a Poisson distribution and the service times are geometrically distributed, we have
\begin{equation*}
\mbox{var}\left( A_z\left( k \right) \right)={\lambda}_z \quad\mbox{ and }\quad \mbox{var}\left( S \right)=\tfrac{1-\mu}{{\mu}^2} \mbox{.}
\end{equation*}
As a result, \eqref{eq:conditional expected sojourn time} reduces to
\begin{equation}
\label{eq:conditional expected sojourn time for mm1}
\mathbb{E}[D | Z=z]=\frac{2\mu - 2{\lambda}_z \mu + \lambda_z^2}{2{\left( \mu - {\lambda}_z \right)}^2} \mbox{.}
\end{equation}
The probability that an arriving task has a priority greater than $z$ is
\begin{align*}
p_z&=\int_{z}^{\infty} \frac{1}{\sqrt{2\pi}} \exp\left({\frac{-u^2}{2}}\right) \, du\\
&=Q(z)\mbox{.}
\end{align*}
Thus, $\lambda_z=\lambda Q(z)$. Substituting this in \eqref{eq:conditional expected sojourn time for mm1} yields:
\begin{equation}
\label{eq:final conditional expected sojourn time for mm1}
\mathbb{E}[D | Z=z]=\frac{2\mu - 2 \mu \lambda Q(z) + \lambda^2 Q^2 (z)}{2{\left( \mu - \lambda Q (z) \right)}^2}\mbox{.}
\end{equation}
Substituting \eqref{eq:conditional expected x for mm1} and \eqref{eq:final conditional expected sojourn time for mm1} into \eqref{eq:iterated cost}, we get the cost function as
\begin{align}
C_o=\rho_{x,z} \int_{z=-\infty}^{\infty} \left( \tfrac{1}{\sqrt{2\pi}} \exp \left({\tfrac{-z^2}{2}}\right) \right) \left( z \frac{2\mu - 2 \mu \lambda Q(z) + \lambda^2 Q^2(z)}{2{\left( \mu - \lambda Q \left( z \right) \right)}^2} \right) dz \mbox{.}
\label{eq:finalcost}
\end{align}
The integral in \eqref{eq:finalcost} is the cost to the principal when the agent is perfectly aligned. For all values of $\mu$ and $\lambda$ (satisfying $\lambda < \mu$), the integral
gives a constant negative value. This is intuitively reasonable, since the absolute values of high priorities
and low priorities are the same and the sojourn times for high priority tasks is smaller than the sojourn times for low priority tasks.
The cost function is completely characterized by $\rho_{x,z}$, up to a multiplicative constant.

\subsubsection{Plotted Results}
\begin{figure}
  \centering
  \includegraphics[scale=0.8]{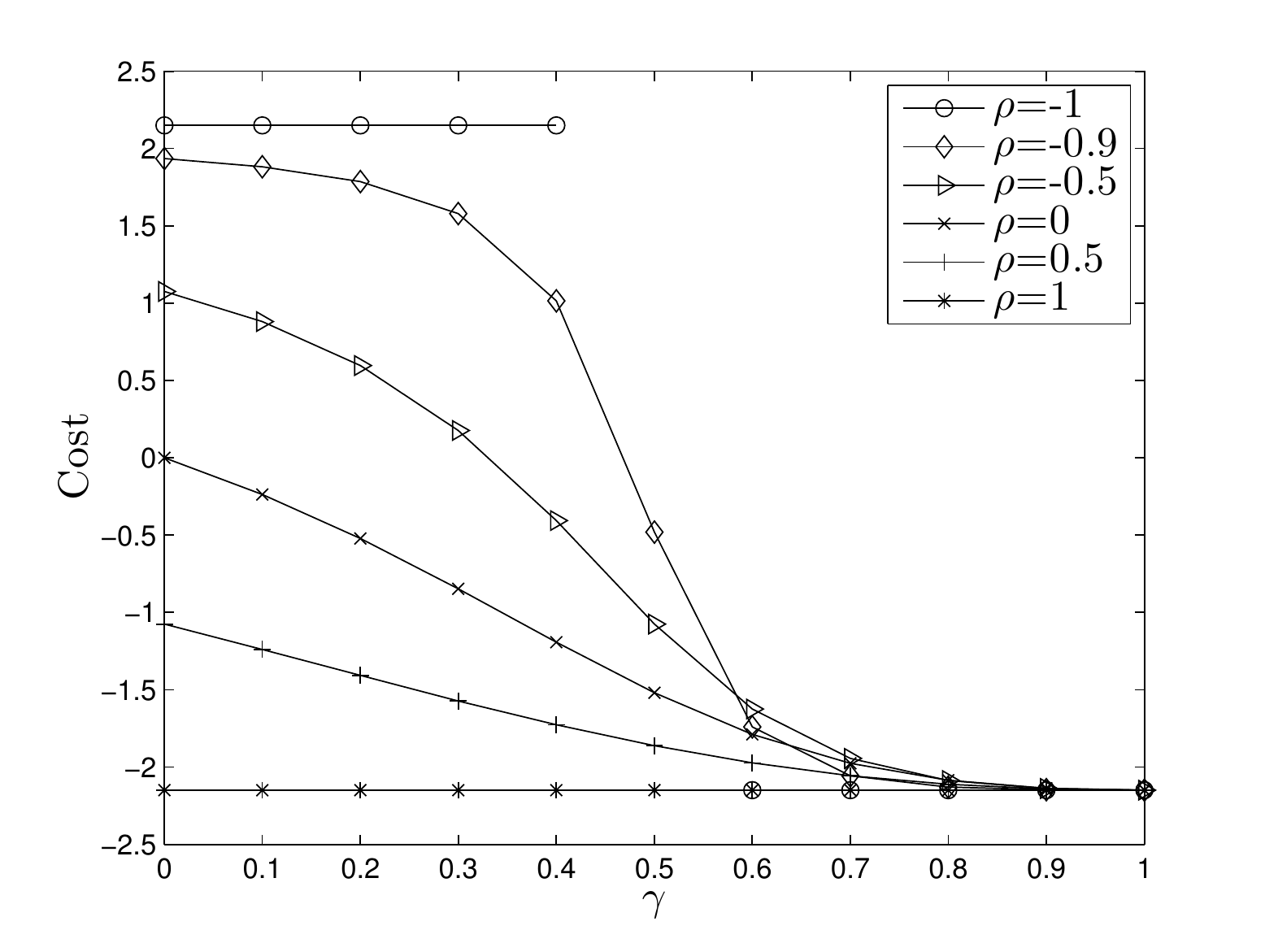}
  \caption{Cost as a function of $\gamma$ for constant values of $\rho$.}
  \label{fig:cost vs gamma}
\end{figure}

\begin{figure}
  \centering
  \includegraphics[scale=0.8]{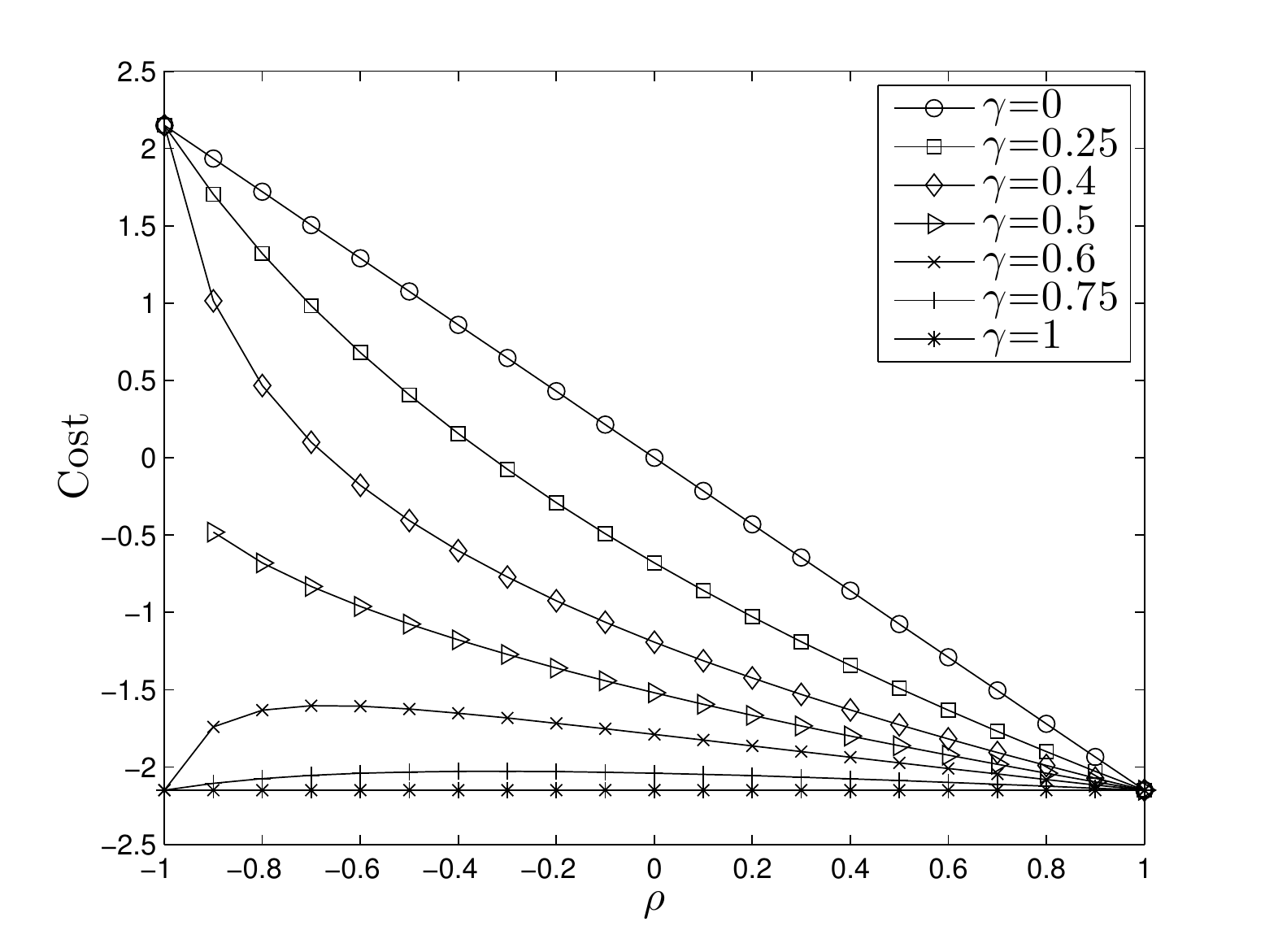}
  \caption{Cost as a function of $\rho$ for constant values of $\gamma$.}
  \label{fig:cost vs rho}
\end{figure}

Continuing with the example, Fig.~\ref{fig:cost vs gamma} plots cost as a function of $\gamma$ for constant values of $\rho$, whereas Fig.~\ref{fig:cost vs rho} plots cost as a function of $\rho$ for constant values of $\gamma$. The plots are drawn for $\mu=0.6$ and $\lambda=0.4$.
Some observations from the plots are as follows (the last one is surprising).
\begin{itemize}
\item For a constant $\rho \in \left( -1,1 \right)$, cost is a monotonically decreasing function
of $\gamma$ (Fig.~\ref{fig:cost vs gamma}). This is expected, since a higher weight given to the
importance of tasks in prioritizing implies less delay for more important tasks, which results in
a smaller cost value.
\item For $\rho=1$, there is complete correlation between the agent's and the principal's interests (Fig.~\ref{fig:cost vs gamma}). So $\gamma$ does not have any effect on the service discipline and hence cost has the minimum value for all $\gamma$. For $\rho=-1$, the agent's and the principal's interests are in complete negative correlation $(Y=-X)$. In this case, $\rho_{x,z}$ becomes
\begin{equation*}
\rho_{x,z}=\frac{2\gamma -1}{| 2\gamma -1 |}.
\end{equation*}
So the cost function attains the maximum value for all $\gamma \in [ 0,0.5)$, and the minimum
value for all $\gamma \in ( 0.5,1 ]$, with a discontinuity at $\gamma=0.5$. At $\gamma=0.5$, the queue operates without priorities.
\item For $\gamma \in ( 0,0.5 )$, cost is a monotonically decreasing function of $\rho$
(Fig.~\ref{fig:cost vs rho}). This is intuitively reasonable, since for a given $\gamma$ if
the agent's interests are more aligned with the principal's interests, cost will decrease.
\item The surprising result is seen in Fig.~\ref{fig:cost vs rho} for $\gamma \in ( 0.5,1 )$. Here, the
cost starts from the minimum value, increases with $\rho$, attains a maximum, and then decreases
to the same minimum value. This implies it is better to have a
completely misaligned agent rather than a slightly aligned agent, if the agent gives more
weight to the importance of the task: \emph{discipline can override lack of alignment}.
\end{itemize}
This last observation can be explained as follows: when $\rho=-1$, $X$ and $Y$ completely determine each
other (misaligned). Thus $Z$ is a completely deterministic function of $X$, and hence $\rho_{x,z}=1$.
With an increase in the value of $\rho$, $X$ and $Y$ are not completely determined by each other;
for a fixed value of $X$, $Y$ can take any value from its support set with a non-zero probability.
Thus, $Z$ is no longer a deterministic function of $X$. Hence, $\rho_{x,z}$ decreases and
consequently cost increases.


These results may help the principal decide how to be maximally effective under some constraints.
For example, the principal can offer some exogenous incentive to the agent so that the agent works in her
favor, and try to tradeoff between the incentive offered and the loss incurred. Later in this paper, we extend this model to accommodate more than one agent and see how it improves the cost for the principal.

Next we look at the setting where the agent prioritizes according to the importance
of the tasks, but performs certain tasks faster than others due to interest.

\subsection{Service rate variation}

Instead of prioritizing by $Z$, the agent now picks a task on the basis of its importance (a
task with a higher $X$ will be executed first). This is a particular case of the previous model
when $\gamma=1$. But now the rate at which a task is executed is an increasing function of
the agent's interest in it, i.e.\
\begin{equation*}
\mu=\mu(Y)\mbox{.}
\end{equation*}
Thus, the agent's priorities coincide exactly with those of the principal, but how fast the agent works depends on his own interest in the task at hand.

The cost function is now
\begin{equation}
\label{eq:cost for variable service rate}
\mathbb{E}[XD]=\mathbb{E}\left[ \mathbb{E}[XD \mid X ] \right)=\mathbb{E}\left[X \mathbb{E}[ D \mid X ] \right].
\end{equation}
This boils down to finding the average sojourn time for a given task. In \cite{WalraevensSB2008},
the expression for $\mathbb{E}[ D | X ]$ is obtained under the assumption that service rates of all tasks in one class are the same. In our model, $\mu(y)$ is the service rate of a task with the agent's interest value $Y=y$. Since $X$ and $Y$ are independent, the service rate of a task with priority $X=x$ is the unconditional service rate of the task, which is the same for each task, given by
\begin{equation}
\label{eq:unconditional service time}
\mathbb{P} \left( S=n \right)=\int \mathbb{P}({S}=n \mid Y=y)f_Y(y) \, dy \, \mbox{.}
\end{equation}
Thus we can use the expression for $\mathbb{E}[D|Z]$ as given in
\eqref{eq:conditional expected sojourn time} (with $Z$ replaced by $X$). Substituting this expression in \eqref{eq:cost for variable service rate}, the cost function is
\begin{align}
\label{eq:cost for one-agent deceitful}
C_{o}=\int \limits_x x f_X(x) dx \left(\frac{\left( 2\mu -{\lambda}_{x}\right) \mbox{var}\left( A_x(k) \right)}{2{\lambda}_{x}{\left( \mu - {\lambda}_{x} \right)}^{2}} 
+ \frac{{\lambda}_{x} {\mu}^{2} \mbox{var}(S)}{2{\left( \mu - {\lambda}_{x} \right)}^{2}} - \frac{{\lambda}_{x}}{2 \left( \mu - {\lambda}_{x} \right)}\right).
\end{align}
Let $\widehat{\mu}$ be the unconditional service rate
and $\widehat{\mbox{var}}(S)$ be the unconditional variance of the service time. Let $\widehat{C_o}$ be the cost for this case.
Once again we look at an example to see the behavior of the cost function. 

\subsubsection{Example}
Assume the arrival process is Poisson and the service time distribution is conditionally geometric, i.e.\
\begin{equation}
\label{eq:conditional service time}
\mathbb{P}({S}=n \mid Y=y)=(1-\mu(y))^{n-1} \mu(y) \mbox{.}
\end{equation}
Clearly, the unconditional service
time distribution is no longer geometric. Since we are interested only in comparing the cost obtained here with the one obtained
with a constant service rate to see the effect of variation in service rate, we assume
$X$ and $Y$ as independent, uniformly distributed random variables in $[0,1]$. 

Also, inspired by the classical
result in psychology called Fitts' Law (in its so-called Shannon formulation) \cite{Fitts1954}, let
\begin{equation}
\label{eq:variable service rate structure}
\mu=\mu_0+\log(1+Y).
\end{equation}
The capacity of a human increases with the demands of the
task in a concave fashion, since human capacity is limited \cite{Kahneman1973}. The choice of
$\log$ ensures that the service rate increases with $Y$, but the rate of increase is a decreasing
function of $Y$; the constant ensures that the service rate does not fall below a minimum value
for any task. Substituting this expression for $\mu$ in \eqref{eq:conditional service time} and
using \eqref{eq:unconditional service time} yields the unconditional probability mass function
for the service time as:
\[
\mathbb{P}({S}=n)= \int_{y=0}^{1} {\left( 1-\mu_0-\log(1+y) \right)}^{n-1} \left( \mu_0+\log(1+y) \right) dy \mbox{.}
\]
To obtain numerical values, we use $\mu_0 = \frac{1}{5}$. This gives
\begin{equation*}
\widehat{\mu}= 0.5 \mbox{ and }\widehat{\mbox{var}}(S) = 3.35 \mbox{.}
\end{equation*}
For Poisson arrivals, $\mbox{var}(\Lambda_x)=\lambda_x$. As $X$ is a uniform random variable,
\begin{equation*}
{\lambda}_x=\lambda p=\lambda \int_{u=x}^{1} f_{U}\left( u \right) du=\lambda \left( 1-x \right) \mbox{.}
\end{equation*}
Substituting the values of ${\lambda}_x$, $\widehat{\mu}$ and
$\widehat{\mbox{var}}(S)$ (in ${\lambda}_x$, $\mu$ and var$(S)$, respectively) in \eqref{eq:cost for one-agent deceitful}, we obtain the following expression for $\widehat{C_o}$ as a function of $\lambda$:
\begin{equation}
\label{eq:variable cost}
\widehat{C_o}=\frac{2\lambda (20\lambda-87)+3 (9\lambda-29)\log(1-2\lambda)}{160{\lambda}^2}.
\end{equation}

Keeping the load constant $\left( \mu=\widehat{\mu} \right)$ and substituting the value of ${\lambda}_x$, we obtain, using \eqref{eq:cost for variable service rate},  an expression for $C_o$ (cost for the case of constant service rate) as a function of $\lambda$:
\begin{equation}
\label{eq:constant cost}
C_o=\frac{{2\lambda}^2-6\lambda-3\log(1-2\lambda)}{8{\lambda}^2}.
\end{equation}
For this case, $\mbox{var}(S)=2$.

\subsubsection{Plotted Results}
\begin{figure}
  \centering
  \includegraphics[scale=0.8]{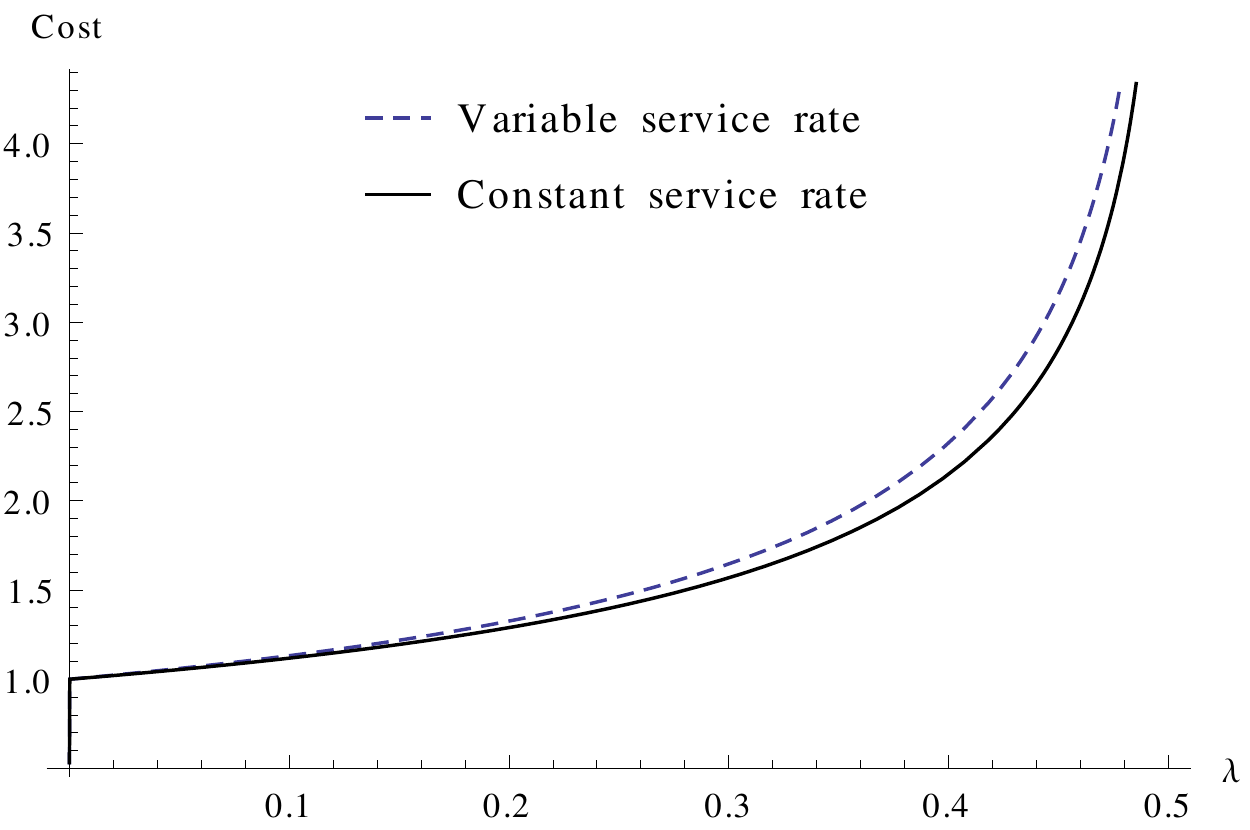}
  \caption{Cost as a function of $\lambda$.}
  \label{fig:cost vs lambda for variable service rate}
\end{figure}

Continuing with the example, Fig.~\ref{fig:cost vs lambda for variable service rate} plots the variation of cost with
the arrival rate. Comparing $\widehat{C_o}$ with $C_o$, it turns out that $\widehat{C_o}>C_o$ for all values
of $\lambda$. This suggests that the cost incurred by the principal is higher when the agent's working
rate depends on his interest in the task, compared to the cost when the agent's working rate is the same
for each task he faces. An explanation for this observation is that the variance of the service time distribution
is greater for the case of variable service rate compared to that for the constant service rate.
Larger variance in the service times typically implies longer expected sojourn times, which leads to a larger cost.

The above observation also points to a larger principle in sociotechnical systems: unlike machines, people have interests and preferences for what they are doing. This makes them more variable and hence less efficient, compared to a machine that works at the same average rate.

Having seen how the agent's self-interests in the tasks and misalignment with the principal affect the principal in the form of cost, we next look at what the principal can do.

\subsection{Principal's role: Incentivizing agent}
How can the principal compensate for the agent's lack of alignment?  Recall that, in the priority-variation setting, the agent prioritizes according to random variable $Z$ given by \eqref{eq:prioritization function}, where $\gamma \in [0,1]$ is the importance the agent assigns to the principal's interests due to some extrinsic motivation offered.  Let $\beta$ be the incentive that the principal gives to the agent, such that the prioritization weight is an increasing function of incentive, $\gamma=f(\beta)$. We aim to find structural properties for the incentive that optimizes the principal's utility. 

The principal's utility function is:
\[
U = -\beta -\theta C(f(\beta))\mbox{,}
\]
where $C(f(\beta))$ is the cost function \eqref{eq:finalcost} for $\gamma=f(\beta)$, and $\theta$ is the constant loss to the principal per unit cost. The utility function comprises the incentive (in monetary value) given to the agent and the loss (in monetary value) incurred due to cost $C$. To find the optimal $\beta$, we solve the following optimization:
\begin{align*}
\max_{\beta} \ & U\\
\mbox{s. t.} \ & \beta \geq 0\mbox{.}
\end{align*}
The solution depends on the function $f$, and the problem may be non-convex optimization in general, but we can easily find numerical solutions. An interesting property of the optimal incentive $\beta$, denoted $\beta^*$, is as follows.
\begin{prop}
\label{prop1}
For any increasing function $\gamma = f(\beta)$, the optimal incentive $\beta^*$ is a non-increasing function of alignment $\rho$.
\end{prop}
\begin{IEEEproof}
See Appendix~\ref{Proof of proposition 1}.
\end{IEEEproof}
We see that if the alignment between the principal and the agent increases, the principal should not increase the payment to the agent.

\section{Multi-agent principal-agent communication}
It seems reasonable to presume the principal's cost should decrease if the principal chooses to allocate a given task to one among several agents,
but can the principal take advantage of the agents' abilities to further reduce her cost?

Consider a principal-agent communication problem where the principal has two agents to allocate tasks to. We assume each task has three kinds of interests: $X$, $Y_1$, and $Y_2$, corresponding to the principal and the two agents, respectively. The principal routes tasks to agents based on limited information about $X$, $Y_1$, and $Y_2$. Agent $i$ prioritizes tasks as some function of $X$ and $Y_i$. As stated in Sec.~\ref{sec:model}, we consider a memoryless task allocation policy to the agents. The cost is denoted $C_m$.

We devise two variations of this model, on similar lines as in the single-agent case.
\begin{itemize}
\item Priority variation with agent's interests: Each agent serves tasks at the same rate, but prioritizes the tasks as a function of the principal's and his own interests. For concreteness, we make the same modeling assumption as in the one-agent priority variation: $X$ and $Y_i, i=1,2$ are distributed as jointly Gaussian random variables.
\item Service rate variation with agent's interests: Each agent prioritizes the tasks according to the principal's interests, but serves tasks at different rates based on his own interests in the tasks. We use the same modeling assumption as in the one-agent case: $X$ and $Y_i, i=1,2$ are distributed as uniform random variables, while $Y_1$ and $Y_2$ are independent random variables.
\end{itemize}
Our first contribution in this section lies in showing that the routing policy that minimizes the cost of certain important tasks, with the overall cost constrained to be minimal, is a threshold policy. Second, we show that diversity of interests in a workforce is beneficial only if the principal can identify it.

\subsection{Priority variation}
Consider $X$ and $Y_i$ as jointly Gaussian random variables with correlation coefficients $\rho_i$, $i=1,2$. We assume that the two agents' interests are conditionally independent. The correlation coefficients capture the principal's alignment with the two agents. We also assume that the means of the three random variables are large positive values, so that almost all tasks have positive priorities. To understand the significance of this assumption, consider the following: Suppose the three random variables are zero-mean. For any queue, given a value of $\lambda_i$ and a well-aligned agent, the cost is negative. Decreasing $\lambda_i$ for a queue then results in an increase in the cost for that queue, since the delay for the low-priority (negative-priority) tasks decreases. This undesirable effect is due to the fact zero-mean random variables have an equal support set on the positive and negative real axis. Note that we did not have to worry about this assumption in the one-agent case because we were dealing with constant $\lambda$.

While this model is an extension of the single-agent priority-variation model, for simplicity let us suppose that agents prioritize tasks strictly according to their own interest and do not care about the principal's interest. Unlike in the one-agent model, the principal routes tasks to the agents based on the information she has about the agents' interests in the tasks, along with her own interest in the tasks. We assume the principal only has statistical information about the agents' interests, in the form of the agents' correlations with the principal, $\rho_1$ and $\rho_2$, rather than information about specific realizations. 

Let $\{R=i\}$ denote the event that a task is routed to the $i$th agent, $i=1,2$. The cost function for two agents can be expanded as follows:
\begin{align}
\label{eq:C_m expanded}
C_m&=\mathbb{E}[XD] \notag \\
&=\mathbb{E}[XD\mathbf{1}_{\{R=1\}}] + \mathbb{E}[ XD\mathbf{1}_{\{R=2\}}] \notag \\
&= \mathbb{E}[XD \mid R=1] \mathbb{P}(R=1) + \mathbb{E}[XD \mid R=2 ] \mathbb{P}(R=2) \mbox{,}
\end{align}
where $\mathbf{1}_{\{ \cdot \}}$ is an indicator random variable, $\mathbb{E}[ XD \mid R=i]$ is the cost for tasks in the $i$th queue, and $\mathbb{P}(R=i)$ is the unconditional routing probability to the $i$th queue.

Note that the joint distribution of $X$ and $D$ given the unconditional routing event $\{ R=i \}$ is the same as the unconditional joint distribution. To see how, since $\mathbb{P}(R=i)$ is the unconditional routing probability, it is the same for all tasks and hence is independent of $X$ and $D$. As a result, $\mathbb{E}[XD \mid R=i]$ is the same as the cost for the single-agent system. Using \eqref{eq:complete cost}, for Poisson distribution of task arrivals and geometric distribution of service rates, $\mathbb{E}[XD \mid R=1]$ is given by
\begin{align}
\mathbb{E}[XD \mid R=1]=\int \limits_y \left( \mu_x + \rho_{1}\frac{\sigma_x}{\sigma_{1,y}}(y-\mu_{1,y}) \right) \left( \frac{2\mu_1 - 2 \mu_1 \lambda_{1,y} + \lambda_{1,y}^2}{2{\left( \mu_1 - \lambda_{1,y} \right)}^2} \right) f_{Y_1}(y) \, dy \mbox{,}
\label{eq:one-agent cost for first queue}
\end{align}
and similarly,
\begin{align}
\mathbb{E}[XD \mid R=2]=\int \limits_y \left( \mu_x + \rho_{2}\frac{\sigma_x}{\sigma_{2,y}}(y-\mu_{2,y}) \right) \left( \frac{2\mu_2 - 2 \mu_2 \lambda_{2,y} + \lambda_{2,y}^2}{2{\left( \mu_2 - \lambda_{2,y} \right)}^2} \right) f_{Y_2}(y) \, dy \mbox{,}
\label{eq:one-agent cost for second queue}
\end{align}
where $\mu_x$ and $\sigma^2_x$ are the mean and variance of the random variable $X$, $\mu_{i,y}$ and $\sigma^2_{i,y}$ are the mean and variance of the random variable $Y_i, i=1,2$, $\mu_i$ is the service rate for the $i$th agent and $\lambda_{i,y}$ is the average number of task arrivals in the $i$th queue in a time step with priorities higher than $y$.

\subsubsection{Minimizing the total cost}
The principal routes tasks based on her own interests in the tasks, $X$, and the statistical information about the agents' interests, $\rho_1$ and $\rho_2$. Let $p(x,\rho_1,\rho_2)$ and $1-p(x,\rho_1,\rho_2)$ be the \emph{conditional} routing probabilities that a task with priority $X=x$ is routed to the first and the second agent, respectively. This can be written as
\begin{align}
p(x,\rho_1,\rho_2)&=\mathbb{P}(\mathbf{1}_{(R=1)}=1 \mid X=x)\\
&=\mathbb{P}(R=1 \mid X=x),
\end{align}
where $\mathbf{1}_{\{R=i \}}$ is $1$ for tasks routed to the $i$th queue and zero otherwise. Therefore, the unconditional routing probability is
\begin{equation}
\label{eq:unconditional routing probability}
\mathbb{P}(R=1)=\mathbb{E}[p(X,\rho_1,\rho_2)].
\end{equation}
Our aim is to find the optimal routing function that minimizes the cost:
\begin{align}
\label{eq:optimization for all tasks}
\min_{p(x, \rho_1, \rho_2)} & C_m \notag \\
\mbox{s. t. } & 0 \leq p(x, \rho_1, \rho_2) \leq 1 \ \ \mbox{ for all } x \in \mathbb{R}.
\end{align}

Before we solve optimization problem \eqref{eq:optimization for all tasks}, the following lemma illustrates the dependency of the cost function on the conditional routing probability.
\begin{lem}
\label{lem1}
The cost for the principal $C_m$ depends on $p(X,\rho_1,\rho_2)$ only through $\mathbb{E}[p(X,\rho_1,\rho_2)]$, the unconditional routing probability.
\end{lem}
\begin{IEEEproof}
See Appendix \ref{Proof of lemma 1}.
\end{IEEEproof}
This shows that the cost does not change with the task allocation function as long as its expected value is fixed.

The minimization in \eqref{eq:optimization for all tasks} can be carried out only over $\mathbb{E}[p(X,\rho_1,\rho_2)]$.
\begin{align}
\label{eq:optimization for all tasks again}
\min_{p(x)} & \ C_m\\
\mbox{s. t. } & 0 \leq \mathbb{E}[p(X,\rho_1,\rho_2)] \leq 1\mbox{.}
\end{align}

Under certain conditions, it can be shown that the above optimization problem is convex; the means of the random variables dictating the convexity of this optimization problem. Refer appendix ~\ref{proof2} for a detailed analysis.

An important point to note here is that the principal routes a task without knowing the realizations of agents' interests (the principal only has statistical knowledge of alignment in the form of $\rho_1$ and $\rho_2$). Therefore the principal needs to avoid overloading any server, and hence the best she can do is maintain an average task routing probability.

Given the optimal unconditional routing probability $\arg\,\min C_m$, the principal can vary the routing probability $p(x, \rho_1, \rho_2)$ to affect the cost for a subset of tasks, keeping the overall cost at the minimum value. An interesting problem then is to determine how the principal chooses the routing probability $p(x, \rho_1, \rho_2)$ to minimize the cost of some ``important'' tasks, keeping overall cost at minimum. Therefore, we next look at finding the optimal routing policy to minimize the cost of certain high-priority tasks. 

\subsubsection{Minimizing the cost for high-priority tasks}
Let $p^*$ be the average routing probability that minimizes the cost in the previous case, i.e.,
\[ 
p^* = \arg\,\min C_m\mbox{.} 
\]
Constraining the average routing probability to $p^*$, we want to find the routing function that minimizes the cost of tasks with priorities higher than a given value $x^*$:
\begin{align}
\label{eq:optimization for high priority tasks}
\min_{p(x)} & \ \mathbb{E}[XD \mid X>x^*] \notag \\
\mbox{ s. t. } & 0 \leq p(x, \rho_1, \rho_2) \leq 1 \ \  \mbox{ for all } x \in \mathbb{R} \notag  \\
& \mathbb{E}[ p(X, \rho_1, \rho_2)]=p^*\mbox{.}
\end{align}

The optimal policy has a threshold structure.
\begin{prop}
\label{prop3}
The optimal routing policy that satisfies \eqref{eq:optimization for high priority tasks} is a threshold policy, given by
\[ p(x, \rho_1, \rho_2)=0, \ \mbox{for }E_{D,1}(x)>E_{D,2}(x), \]
\[ p(x, \rho_1, \rho_2)=1, \ \mbox{for }E_{D,1}(x)<E_{D,2}(x), \]
\[ p(x, \rho_1, \rho_2) \in [0,1], \ \mbox{for }E_{D,1}(x)=E_{D,2}(x), \]
where $x > x^*$ and $E_{D,i}(x)$ is the average delay of a task with priority $x$ given it is routed to the $i$th queue.
\end{prop}
\begin{IEEEproof}
See Appendix~\ref{app:prop3}.
\end{IEEEproof}
This is the optimal threshold policy for tasks with priorities greater than $x^*$. To find the complete routing function, we use the constraint $\mathbb{E}[p(X)]=p^*$.
Further, it is shown in Appendix~\ref{app:prop3} that $E_{D,1}(x)$ and $E_{D,2}(x)$ depend on $p(x, \rho_1, \rho_2)$ only through $\mathbb{E}[p(X, \rho_1, \rho_2)]$.

This shows that by keeping the average routing probability fixed, the principal can vary the routing function so as to vary the cost of a particular subset of tasks, without affecting the overall cost. This turns out to be an added advantage of having two agents, along with the obvious benefit of less cost.

Next, we consider the service rate variation version of the model: agents prioritize tasks according to the principal's interests, but vary their service rates according to their own interests.

\subsection{Service rate variation}
An agent's service rate is a function of interest in the task, i.e.,
\[
\mu=\mu(Y_i)\mbox{.}
\]
Since the working rate of a person cannot increase as fast as his interest, we assume $\mu(Y)$ is a concave function. As in \eqref{eq:variable service rate structure}, we assume the following functional form for $\mu(Y_i)$:
\[ 
\mu(Y_i)=\mu_0 + \log(1+Y_i)\mbox{.}
\]

We aim to characterize the impact of diversity in agent interests on the cost to the principal, analyzing performance when the principal has complete knowledge of the agents' interest-realizations. In this setting, we look at several cases based on the correlation between the agents' interests, comparing cost against agents working at a constant rate. Later we consider a more practical setting where the principal has noisy measurements of the agents' interest-realizations.
Similar to single-agent service-rate variation, here we assume $X, Y_1, Y_2$ are marginally uniformly distributed in $[0,1],$ and that $X$ is independent of $Y_i, i=1,2$.

To find the complete cost function, recall from \eqref{eq:C_m expanded} that the cost function for the two-agent model is:
\begin{align}
\label{eq: C_m expanded again}
C_m = \mathbb{E}[XD \mid R=1] \mathbb{P}(R=1) + \mathbb{E}[XD \mid R=2 ] \mathbb{P}(R=2) \mbox{,}
\end{align}
where $\mathbb{E}[XD \mid R=i]$ is the cost for the tasks routed to the $i$th queue and $\mathbb{P}(R=i)$ is the unconditional routing probability to the $i$th queue. Once again we see that the joint distribution of $X$ and $D$ given the unconditional routing function $(R=1)$ is the same as the unconditional joint distribution, since $(R=1)$ is independent of both $X$ and $D$. 
Using \eqref{eq:cost for one-agent deceitful}, for Poisson distribution of task arrival process and geometric distribution of conditional service times, $\mathbb{E}[XD \mid R=1]$ can be expanded as
\begin{align}
\label{eq:cost for one-agent deceitful agent one}
\mathbb{E}[XD \mid R=1]=\int \limits_x x f_X(x) dx \left(\frac{\left( 2\mu_1 -{\lambda}_{1,x}\right)}{2{\left( \mu_1 - {\lambda}_{1,x} \right)}^{2}} 
+ \frac{{\lambda}_{1,x} \mu_1^2 \mbox{var} \left( S_1 \right)}{2{\left( \mu_1 - {\lambda}_{1,x} \right)}^{2}} - \frac{{\lambda}_{1,x}}{2 \left( \mu_1 - {\lambda}_{1,x} \right)}\right)\mbox{,}
\end{align}
and for the second agent as
\begin{align}
\label{eq:cost for one-agent deceitful agent two}
\mathbb{E}[XD \mid R=2]=\int \limits_x x f_X(x) dx \left(\frac{\left( 2\mu_2 -{\lambda}_{2,x}\right)}{2{\left( \mu_2 - {\lambda}_{2,x} \right)}^{2}} 
+ \frac{{\lambda}_{2,x} \mu_2^2 \mbox{var} \left( S_2 \right)}{2{\left( \mu_2 - {\lambda}_{2,x} \right)}^{2}} - \frac{{\lambda}_{2,x}}{2 \left( \mu_2 - {\lambda}_{2,x} \right)}\right)\mbox{.}
\end{align}
where $\mu_i$ is the service rate of the $i$th agent, $\lambda_{i,x}$ is the average number of task arrivals in the $i$th queue in a time step with priorities higher than $x$, and var$(S_i)$ is the variance of execution times for the $i$th agent.

As the tasks are assigned to the two agents, the distribution of the agent priorities in the queues would be different from their unconditional distributions. Therefore, we denote by random variables $Z_1$ and $Z_2$ the agent priorities that reach the first queue and the second queue, respectively. In other words, $Z_1$ and $Z_2$ represent the agent priorities of the tasks in the first and the second queue, respectively. Note that the $X$ values remain the same. To find the cost function for different cases, we first find the distribution of $Z_1$ and $Z_2$.
Then we find the values for the unconditional service rates, $\mu_1$ and $\mu_2$, and the variance of the service times, var$(S_1)$ and var$(S_2)$, using the following set of equations:
\[ \mathbb{P}(S_i=n)=\mathbb{E}\left[ \mathbb{P}(S_i=n \mid Z_i) \right], \ i=1,2. \]
\begin{equation}
\label{eq:conditional service time is geometric}
\mathbb{P}(S_i=n \mid Z_i=z)={\left( 1-\mu(z) \right)}^{n-1} \mu(z), \ i=1,2.
\end{equation}
\[ \mu(Z_i)=\mu_0 + \log (1+Z_i), \ i=1,2. \]
where \eqref{eq:conditional service time is geometric} follows from the  assumption that the conditional service time of a task is geometrically distributed. We then substitute these values in the expression for cost and compare different cost functions by plotting. To obtain numerical values, we use $\mu_0=\frac{1}{5}$.

\subsubsection{Effect of diversity}
We assume that for each task, the principal knows the exact realizations of the agents' interests. We assume the following routing policy for the principal: Each task is routed to the agent who has a higher priority for that task. Then, since $Y_1$ and $Y_2$ are identically distributed, we have
\[ \mathbb{P}(R=1)=\mathbb{P}(R=2)=\frac{1}{2}\mbox{.} \]
Clearly, the routing policy depends on the relation between the agents' interests. Let us first express $Z_1$ and $Z_2$ in terms of $Y_1$ and $Y_2$:
\begin{align}
\label{Z_1 in terms of Y_1}
Z_1 = \begin{cases} Y_1, & \mbox{when } Y_1 > Y_2, \quad \mbox{w.p. }1 \\
Y_1, & \mbox{when } Y_1 < Y_2, \quad \mbox{w.p. }0. \end{cases}
\end{align}
\begin{align}
\label{Z_2 in terms of Y_2}
Z_2 = \begin{cases} Y_2, & \mbox{when } Y_2 > Y_1, \quad \mbox{w.p. }1 \\
Y_2, & \mbox{when } Y_2 < Y_1, \quad \mbox{w.p. }0. \end{cases}
\end{align}
Let $\rho_G$ denote the correlation between the two agents.
We analyze this regime for different levels of diversity in the agents' interests.
\begin{itemize}
\item[A)] \textit{Perfect diversity}  $(\rho_G=-1)$: The two agents' interests are completely negatively correlated, \textit{i.e.}, $Y_2=1-Y_1$. Using \eqref{Z_1 in terms of Y_1} and \eqref{Z_2 in terms of Y_2},
\begin{align*}
\mathbb{P}(Z_1 \leq z) &= \mathbb{P}(Y_1 \leq z \mid Y_1 > Y_2)\\
&= \mathbb{P}(Y_1 \leq z, Y_1 > \tfrac{1}{2}) / \mathbb{P}(Y_1 > \tfrac{1}{2})\\
&= 2z-1\mbox{.}
\end{align*}
Similarly, $\mathbb{P}(Z_2 \leq z)=2z-1$.
As a result, $Z_1, Z_2 \sim \mathcal{U}[\frac{1}{2},1]$. Once again, computations yield:
\[ \mu_i=0.75, \quad \text{var}(S_i)=0.5, \ i=1,2.\]

\item[B)] \emph{Imperfect diversity} $(\rho_G \in (-1,0])$: Consider next a more realistic case, where the diversity in agents' interests is not perfect. In this case, we model $Y_1$ and $Y_2$ as the marginals of a Gaussian copula:
\begin{align}
\label{eq:copula}
C_{\rho_G}(y_1,y_2)=\int \limits_{-\infty}^{{\Phi}^{-1}(y_1)} \int \limits_{-\infty}^{{\Phi}^{-1}(y_2)} \frac{1}{2\pi\sqrt{1-\rho_G^2}} \exp \left( -\frac{s^2 -2\rho_G st + t^2}{2(1-\rho_G^2)} \right) ds \, dt,
\end{align}
where ${\Phi}^{-1}(\cdot)$ is the quasi-inverse of standard normal distribution, and $\rho_G$ is the correlation in the bivariate normal distribution. Due to symmetry, the distribution of $Z_1$ and $Z_2$ will be the same. We compute the service rate and variance of service times for three values of $\rho_G$. Using \eqref{Z_1 in terms of Y_1}, \eqref{Z_2 in terms of Y_2} and \eqref{eq:copula}, in Appendix~\ref{app:copula} we obtain the following values of $\mu$ and var$(S)$:
\begin{align*}
\mu_i&=0.68, \quad \text{var}(S_i)=2.48, \quad \rho_G=-0.8, \\
\mu_i&=0.59, \quad \text{var}(S_i)=2.73, \quad \rho_G=-0.4, \\
\mu_i&=0.53, \quad \text{var}(S_i)=2.99, \quad \rho_G=0.
\end{align*}
\end{itemize}

For comparison, we consider the setting where agents have constant service rate and assume independent Bernoulli routing for each task with parameter $p_0$.  By symmetry, we see the minimum cost will be achieved at $p_0=\frac{1}{2}$ which yields unconditional service times that are geometrically distributed. We choose the average service rates to be equal to the unconditional average service rates for independent agents. Thus, $\mu_1=\mu_2=0.53$, which gives $\text{var}(S_i)=1.67, \ i=1,2$.

\subsubsection*{Plotted results}
\begin{figure}
  \centering
  \includegraphics[scale=0.8]{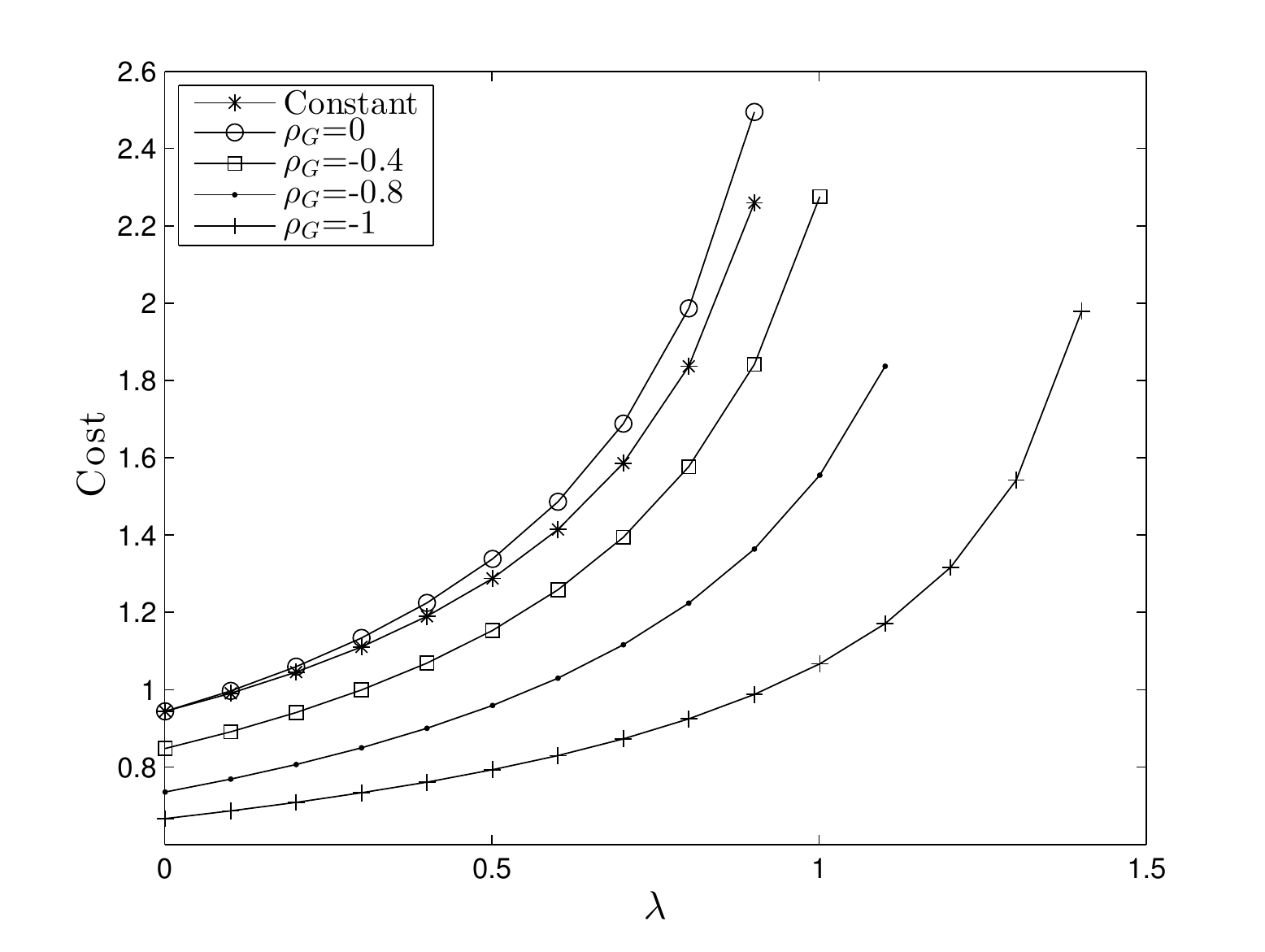}
  \caption{Cost as a function of the arrival rate $\lambda$ to show the effect of diversity among agents' interests: \textit{(a)} Machine case (constant), \textit{(b)} Independent agents ($\rho_G=0$), \textit{(c),(d)} Imperfect negative correlation (${\rho}_{G}=-0.8$, ${\rho}_{G}=-0.4$), \textit{(e)} Completely diverse agents ($\rho_G=-1$)}
\label{fig:cost vs lambda figure 1}
\end{figure}

Fig.~\ref{fig:cost vs lambda figure 1} shows the cost function for different values of correlation between the agent interests, along with the machine (constant service) case, against the arrival rate $\lambda$. The fact variance is lower for the machine case compared to the independent agents case is manifested in the form of less cost for the machine case. This is similar to the single-agent model, where the machine case has less cost than the variable agent case. However, note that the cost is significantly lower than the machine case when there is a high level of interest diversity among the agents. This is since one of the agents is often highly interested in each task, and the principal routes to that agent. Therefore, variable agents are desirable for the principal if they have enough diversity, and the principal has knowledge of the diversity. The principal can take  advantage of diverse agents interests by routing each task appropriately according to agent interests.

The principal, however, may not have complete knowledge of agents' interest realizations. We therefore consider a principal having quantal measurements of the agents' interests.

\subsubsection{Quantal perception of the agents' interest-realizations}
Suppose the principal does not have exact knowledge of the agents' interest-realizations.  When interest-realizations are either high or low, the principal can route them correctly, but for intermediate levels of interest, the principal may perceive erroneously.
Specifically, suppose the principal correctly perceives high-interest values in the range $[q,1]$ and low-interest values in the range $[0,q]$. Since the principal is unsure about interest values in the range $(1-q,q)$, she makes a random routing decision in this case of erasure. Let $p$ be the probability an `unsure' task is routed to the first agent. For computation, we consider $p=\frac{1}{2}$, implying $\mathbb{P}(R=1)=\mathbb{P}(R=2)=\frac{1}{2}$ in \eqref{eq: C_m expanded again}. Let us find the cost function, considering perfect negative correlation among the agent interests.

We first compute the distributions of $Z_1$ and $Z_2$. Observe that $Z_1$ can be written as
\[
Z_1 = \begin{cases} Y_1, & \mbox{when } Y_1 \in [q,1], \quad \mbox{w.p. }1 \\
Y_1, & \mbox{when } Y_1 \in [1-q,q], \quad \mbox{w.p. }\frac{1}{2}. \end{cases}
\]
Therefore, $Z_1$ is uniform in both intervals. After normalizing:
\[
f_{Z_1}(z) = \begin{cases} 2, & z \in [q,1] \\
1, & z \in [1-q,q].\end{cases}
\]
Since, the region of uncertainty is symmetric, and $Y_2=1-Y_1$, the distribution of $Z_2$ is the same as $Z_1$ and the cost function will be equal for both queues. With computation, we can find $\mu$ and var$(S)$, which will be functions of $q$. Therefore, the cost function for this case will be a function of $q$ as well.

What happens as the number of agents increases? The cost obviously decreases with an increase in the number of agents. In addition, as the following proposition shows, the dependence of cost on $\lambda$ vanishes as the number of agents goes to infinity, while the sojourn time of a task approaches its service time.
\begin{prop}
The asymptotic cost $C_{\infty}$ is given by
\begin{equation*}
\label{prop4}
C_{\infty}=\frac{1}{2\mu(y)}\bigg|_{y=1},
\end{equation*}
which implies that each task receives the maximum service rate in the asymptotic case.
\end{prop}
\begin{IEEEproof}
See Appendix~\ref{app:prop4}.
\end{IEEEproof}
As the number of agents increases, the service rate of each task approaches to the maximum possible value. Again, when there is high diversity in the agents' interests, and the principal is aware of it, the principal can obtain higher working rate for each task.

\subsubsection{Plotted results}
\begin{figure}
  \centering
  \includegraphics[scale=0.8]{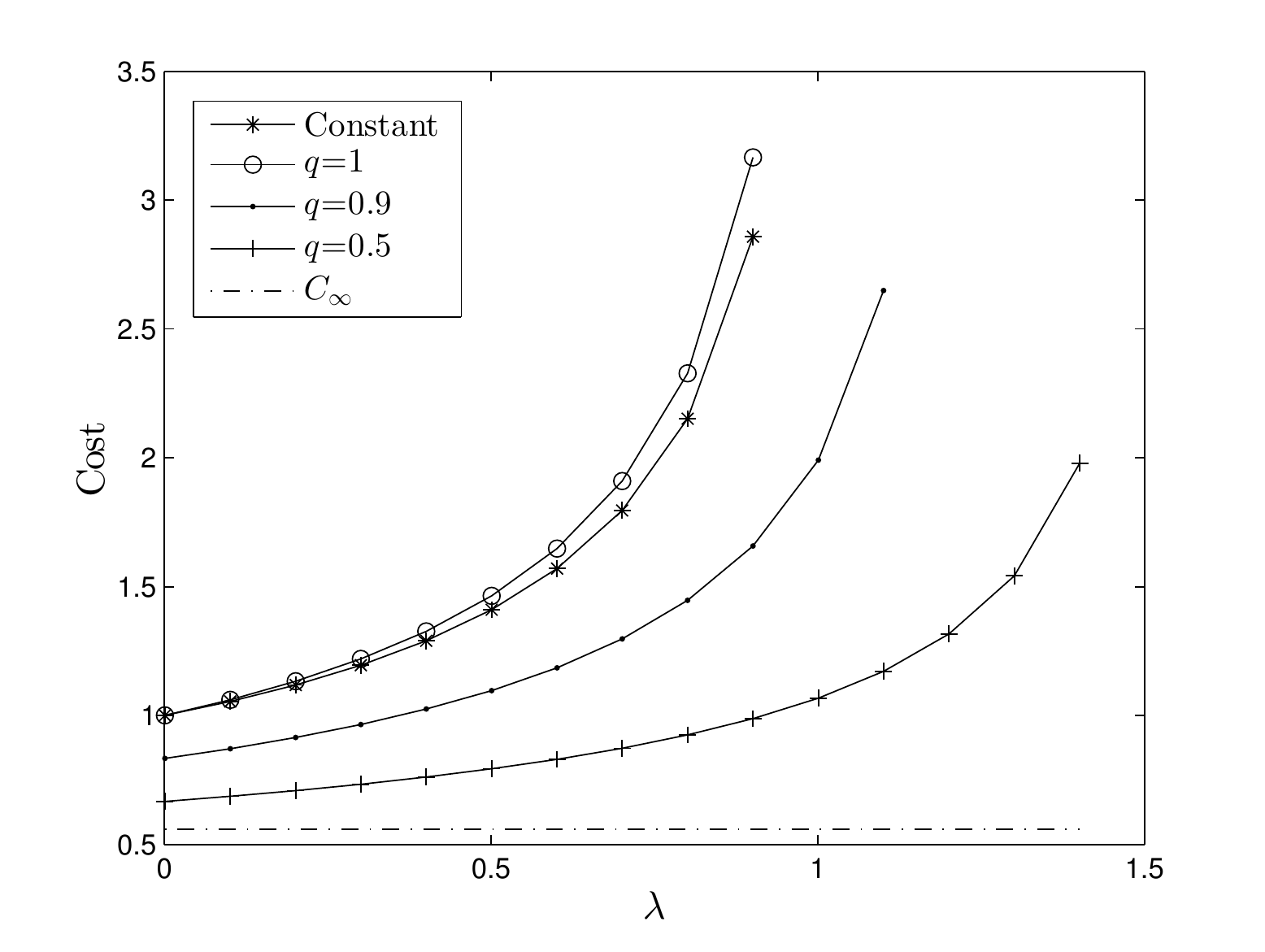}
  \caption{Cost as a function of $\lambda$ for different values of $q$ along with the machine case and the asymptotic case, showing the effect of principal's knowledge about the agents' interest-realizations. Note that $q=1$ and $q=0.5$ correspond to $\rho_G=0$ and $\rho_G=-1$ in Fig. \ref{fig:cost vs lambda figure 1}, respectively.}
  \label{fig:cost vs lambda figure 2}
\end{figure}

Fig.~\ref{fig:cost vs lambda figure 2} plots cost functions for different values of erasure threshold $q$ along with the machine case, for the same $\lambda$. Note that $q=1$ corresponds to random routing due to fully incomplete knowledge of agents' interest-realizations. This is equivalent to the case when the principal has exact knowledge of the agents' interest realizations while the agents' interests are independent. A similar correspondence holds between $q=0.5$ and $\rho_a=-1$.

Evidently, the $q=1$ curve performs worse than the machine case, while $q=0.5$ performs much better. This shows that the diversity in the agents' interests is favorable for the principal if the principal is aware of it, but can be undesirable otherwise. Notice also that the cost function for $q=0.9$ is less than the machine case. This shows that when there is high diversity, the principal needs to route correctly only a few high priority tasks.

\section{Conclusion}
Motivated by the fact that people perform routine tasks by prioritizing them, we viewed the human priority queue from a principal-agent perspective. For the single-agent setting, our model serves to quantify the effect of misalignment between the priorities of the task sender and the receiver. In particular, we characterize how the cost to the sender varies with the correlation between the agent's and the principal's interests, and with the prioritizing function chosen by the agent. We established that while it is favorable for the principal to have an agent with high intrinsic alignment, it is possible for the principal to attract the agent in her favor by means of extrinsic motivation, if the agent is not intrinsically well-aligned with the principal. We also argued that humans typically have a larger variance in task execution compared to machines, due to interests that govern their functioning. Lastly, we obtained an interesting structural property of the incentive that the principal offers to the agent as an extrinsic motivation.

In the two-agent setting, we established that the optimal task assignment policy that minimizes the cost for all the tasks depends only on the average routing probability. In addition, we showed that a task assignment policy can be chosen to affect a subset of the task without changing the overall cost. Finally, we showed that unlike the one-agent case, variability in human interests can be potentially favorable to the principal and proved that a workforce with diverse interests is beneficial if the principal can identify it, but otherwise not.

Validating these qualitative results certainly necessitates justifying robustness of the mathematical model used. As such, we emphasize that while the exact mathematical model may not precisely capture the real-world scenario, the arguments and results we have adopted capture the basic essence of the principal-agent queuing problem in ways that are described in the behavioral and social sciences.

\appendices
\section{Proof of Proposition~\ref{prop1}}
\label{Proof of proposition 1}
Recall Fig.~\ref{fig:cost vs rho}, which relates cost $C$ to alignment $\rho$. Also recall the utility function for the principal is
$U=-\beta-\theta C(f(\beta))$, where $\beta$ is the incentive paid to the agent, $\theta$ is the per unit weighted sojourn time cost, and $C(\cdot)$ is the cost function given in \eqref{eq:finalcost}. $C(\cdot)$ is simply $C(\cdot)=C_1 \rho_{x,z}$, where $C_1$ is a negative constant given by
\begin{align*}
C_1=\int_{z=-\infty}^{\infty} \left( \tfrac{1}{\sqrt{2\pi}} \exp \left({\tfrac{-z^2}{2}}\right) \right) \left( z \frac{2\mu - 2 \mu \lambda Q(z) + \lambda^2 Q^2(z)}{2{\left( \mu - \lambda Q \left( z \right) \right)}^2} \right) dz,
\end{align*}
and $\rho_{x,z}$ is the correlation between the principal's and agent's interests.
Taking $\gamma=f(\beta)$, and using the expression for $\rho_{x,z}$ in \eqref{eq:rho_xy}, the utility function is:
\[ 
U=-\beta - \theta \frac{C_1 \left( \gamma + \rho \left( 1-\gamma \right) \right)}{\sqrt{{\gamma}^2 + {\left(1-\gamma \right)}^2 + 2\rho \gamma \left( 1-\gamma \right)}}\mbox{.}
\]
Let $\beta^*$ be the optimum incentive and $\gamma^*=f(\beta^*)$. Recalling $f$ is an increasing function, we need to show $\beta^*$ is a non-increasing function of $\rho$. Consider the following:
\begin{equation*}
\max_{\beta^*} \, U = -\beta^* - \theta \frac{C_1 \left( \gamma^* + \rho \left( 1-\gamma^* \right) \right)}{\sqrt{{\gamma^*}^2 + {\left(1-\gamma^* \right)}^2 + 2\rho \gamma^* \left( 1-\gamma^* \right)}}\mbox{.}
\end{equation*}
At $\rho=-1$,
\begin{equation*}
\max_{\beta^*} \, U \bigg|_{\rho=-1}=-\beta^* - \theta \frac{C_1 \left( 2\gamma^* -1 \right)}{|2\gamma^*-1|}\mbox{.}
\end{equation*}
Since $C_1$ is negative, 
\[ 
\underset{\gamma^*}{\argmax}\,\,U \bigg|_{\rho=-1}={\left(\frac{1}{2}\right)}^+. 
\]
Now, for all values of $\gamma > \frac{1}{2}$, cost is a decreasing function of $\rho$.
This implies $\underset{\gamma^*}{\argmax}\,\,U$ does not increase with increase in $\rho$. Since $\gamma^*$ is an increasing function of $\beta^*$, the result follows.

\section{Proof of Lemma~\ref{lem1}}
\label{Proof of lemma 1}
From \eqref{eq:C_m expanded}, the cost function for the two-agent case is
\begin{align}
\label{eq:cost for two again}
C_m = \mathbb{E}[XD \mid R=1]\mathbb{P}(R=1) + \mathbb{E}[XD \mid R=2]\mathbb{P}(R=2)\mbox{,}
\end{align}
where $\mathbb{P}(R=i)$ is the unconditional routing probability to the $i$th queue, and $\mathbb{E}[XD | R=i]$ is the cost for tasks in the $i$th queue. We develop the proof for the first queue, and the proof for the second queue will follow identically. Using \eqref{eq:one-agent cost for first queue}, the cost for the tasks in the first queue is
\begin{align}
\label{eq:expression for cost for single queue again}
\mathbb{E}[XD \mid R=1]=\int \limits_y \left( \mu_x + \rho_{1}\frac{\sigma_x}{\sigma_y}(y-\mu_{1,y}) \right) \left( \frac{2\mu_1 - 2 \mu_1 \lambda_{1,y} + \lambda_{1,y}^2}{2{\left( \mu_1 - \lambda_{1,y} \right)}^2} \right) f_{Y_1}(y) \, dy \mbox{,}
\end{align}
where $\mu_x$ and $\mu_{1,y}$ are the means of $X$ and $Y_1$ respectively, $\sigma_x^2$ and $\sigma_y^2$ are the respective variances, $\rho_{1}$ is the correlation coefficient, $\mu_1$ is the unconditional service rate, and $\lambda_{1,y}$ is the average number of task arrivals in a time step with priorities higher than $Y_1=y$. Note that none of these variables except $\lambda_{1,y}$ can depend on $p(x)$. Therefore, let us derive the expression for $\lambda_{1,y}$.

Let $\Lambda_{1,y}(z)$ be the generating function for the number of tasks that arrive to the first queue in a time slot with priorities higher than $Y_1=y$. Denote by $E_y(k)$ the event that $k$ tasks arrive to the first queue in a time slot with priorities higher than $Y_1=y$. Then,
\begin{align}
\label{eq:generating function for the no of tasks to the first agent}
\Lambda_{1,y}(z)=& \sum\limits_{k=0}^{\infty} \mathbb{P}\left(E_y(k)\right){z}^k.
\end{align}
Note that $\mathbb{P}\left(E_y(k)\right)$ can be written as
\begin{align}
\label{eq:E_y(k)}
&\mathbb{P}\left(E_y(k)\right) \notag \\
&= \sum\limits_{n=k}^{\infty} \mathbb{P}\left(E_y(k) \mid E_1(n)\right) \mathbb{P}\left(E_1(n)\right) \notag \\
&= \sum\limits_{n=k}^{\infty} \mathbb{P}\left(E_y(k) \mid E_1(n)\right) \sum\limits_{m=n}^{\infty} \mathbb{P}\left(E_1(n) \mid E(m)\right)\mathbb{P}\left(E(m)\right).
\end{align}
where we define the events $E_1(n)$ and $E(m)$ as: $E_1(n) \triangleq n$ tasks arrive to the first queue in a time slot, and $E(m) \triangleq m$ tasks arrive to the principal in a time slot.

Let $p_y$ be the probability that an arriving task has a priority higher than $Y_1=y$. Then,
\begin{equation}
\label{eq:E_y(k)|E_1(n)}
\mathbb{P}\left(E_y(k) \mid E(n)\right)=\binom{n}{k} {(p_y)}^k {(1-p_y)}^{n-k}.
\end{equation}
Let $a(m)$ be the probability that $m$ tasks arrive to the principal in a time slot. Let $\Lambda(z)$ be the generating function for the number of task arrivals to the principal in a time slot, so that
\[ \Lambda(z)=\sum\limits_{m=0}^{\infty}a(m)z^m. \]
Recall from \eqref{eq:unconditional routing probability} that $\mathbb{E}[p(X,\rho_1,\rho_2)]$ is the unconditional probability of routing a task to the first agent,
which for brevity we denote by $q$.  
\begin{align}
\label{eq:E_1(n)|E(m)}
&\mathbb{P}\left(E_1(n) \mid E(m)\right) \notag \\
&=\binom{m}{n} q^n (1-q)^{m-n}\mbox{.}
\end{align}
Substituting \eqref{eq:E_y(k)}, \eqref{eq:E_y(k)|E_1(n)} and \eqref{eq:E_1(n)|E(m)} in \eqref{eq:generating function for the no of tasks to the first agent} gives
\begin{align*}
&\Lambda_{1,y}(z) \\
&= \sum\limits_{k=0}^{\infty} z^k \sum\limits_{n=k}^{\infty} \binom{n}{k} {(p_y)}^k {(1-p_y)}^{n-k} \sum\limits_{m=n}^{\infty} a(m)\binom{m}{n} q^n (1-q)^{m-n}\\
&= \sum\limits_{m=0}^{\infty} a(m) \sum\limits_{n=0}^{m} \left( \sum\limits_{k=0}^{n} \binom{n}{k} {(p_y)}^k {(1-p_y)}^{n-k} z^k \right)\binom{m}{n} q^n (1-q)^{m-n}\\
&= \sum\limits_{m=0}^{\infty} a(m) \sum\limits_{n=0}^{m} {\left( zp_y + 1 - p_y \right)}^n \binom{m}{n} q^n (1-q)^{m-n} \\
&= \sum\limits_{m=0}^{\infty} a(m) { \left(\left( zp_y + 1 - p_y \right) q\right)}^m \\
&\stackrel{(a)}{=} \Lambda \left(\left( zp_y + 1 - p_y \right) \left( \mathbb{E}\left[p(X,\rho_1,\rho_2)\right] \right)\right),
\end{align*}
where $(a)$ is obtained by substituting back $q=\mathbb{E}\left[p(X,\rho_1,\rho_2)\right]$. As a result,
\begin{equation} 
\lambda_{1,y}=\Lambda'_{1,y}(z)=\lambda p_y \mathbb{E}\left[p(X,\rho_1,\rho_2)\right]
\end{equation}
and similarly
\begin{equation}
\lambda_{2,y}=\Lambda'_{2,y}(z)=\lambda p_y \left( 1-\mathbb{E}\left[p(X,\rho_1,\rho_2)\right] \right)\mbox{.}
\end{equation}
Therefore, the overall cost function depends on $p(x)$ only through $\mathbb{E}\left[p(X,\rho_1,\rho_2)\right]$.

\section{On the convexity of multi-agent cost function}
\begin{prop}
\label{prop2}
$C_m$ is a convex function of $\mathbb{E}[p(X, \rho_1, \rho_2)]$ when $\mu_x - \mu_{y_i}$ is sufficiently large.
\end{prop}
\begin{IEEEproof}
\label{proof2}
Let $q = \mathbb{E}[p(X, \rho_1, \rho_2)]$. Using \eqref{eq:C_m expanded}, we expand $C_m$ as follows
\begin{align}
\label{eq:cost for two once again}
C_m = \mathbb{E}[XD \mid R=1]\mathbb{P}(R=1) + \mathbb{E}[XD \mid R=2]\mathbb{P}(R=2)\mbox{,}
\end{align}
where, as before, $\{ R=i \}$ is the event that a task is routed to the $i$th queue, and $\mathbb{E}[XD \mid R=i]$ is the cost for tasks in the $i$th queue, given by
\begin{align}
\label{eq:expression for cost for single queue once again}
\mathbb{E}[XD \mid R=i]=\int \limits_y \left( \mu_x + \rho_{i}\frac{\sigma_x}{\sigma_{yi}}(y-\mu_{yi}) \right) \left( \frac{2\mu_i - 2 \mu_i \lambda_{yi} + \lambda_{yi}^2}{2{\left( \mu_i - \lambda_{yi} \right)}^2} \right) f_{Y_i}(y) \, dy \mbox{.}
\end{align}
For brevity, and without loss of generalization, let $\sigma_x=\sigma_{y_i}=1$. For convexity, we must show that the second derivative of \eqref{eq:cost for two once again} is non-negative. Substituting \eqref{eq:expression for cost for single queue once again} in \eqref{eq:cost for two once again} and differentiating gives
\begin{align}
\label{eq:second derivative of cost}
\frac{d^2C_m}{dq^2}&= 4\int \limits_y \left( \mu_x + \rho_1(y-\mu_{y_1}) \right) \left( \frac{(2\mu-{\mu}^2)\lambda Q(y-\mu_{y_1})}{2{(\mu-\lambda q Q(y-\mu_{y_1}))}^3} \right) f_{Y_1}(y) dy \notag \\
&+ 6 q \int \limits_y \left( \mu_x + \rho_1(y-\mu_{y_1}) \right) \left( \frac{(2\mu-{\mu}^2){\lambda}^2 Q^2(y-\mu_{y_1})}{2{(\mu-\lambda q Q(y-\mu_{y_1}))}^4} \right) f_{Y_1}(y) dy \notag \\
&+ 4\int \limits_y \left( \mu_x + \rho_2(y-\mu_{y_2}) \right) \left( \frac{(2\mu-{\mu}^2)\lambda Q(y-\mu_{y_2})}{2{(\mu-\lambda (1-q) Q(y-\mu_{y_2}))}^3} \right) f_{Y_2}(y) dy \notag \\
&+ 6(1-q)\int \limits_y \left( \mu_x + \rho_2(y-\mu_{y_2}) \right) \left( \frac{(2\mu-{\mu}^2)\lambda^2 Q^2(y-\mu_{y_2})}{2{(\mu-\lambda (1-q) Q(y-\mu_{y_2}))}^4} \right) f_{Y_2}(y) dy.
\end{align}
Consider the first term:
\begin{equation}
\label{eq:first term}
4\int \limits_y \left( \mu_x + \rho_1(y-\mu_{y_1}) \right) \left( \frac{(2\mu-{\mu}^2)\lambda Q(y-\mu_{y_1})}{2{(\mu-\lambda q Q(y-\mu_{y_1}))}^3} \right)f_{Y_1}(y) \, dy.
\end{equation}
To show that \eqref{eq:first term} is positive, we write  the following inequality:
\begin{align}
\label{eq:first term inequality}
&\int \limits_y \left( \mu_x + \rho_1(y-\mu_{y_1}) \right) \left( \frac{(2\mu-{\mu}^2)\lambda Q(y-\mu_{y_1})}{2{(\mu-\lambda p_0 Q(y-\mu_{y_1}))}^3} \right)f_{Y_1}(y) \, dy \notag \\
& \geq \frac{(2\mu-{\mu}^2)\lambda}{{\mu}^3} \int \limits_y \left( \mu_x + \rho_1(y-\mu_{y_1}) \right) Q(y-\mu_{y_1})f_{Y_1}(y) \, dy.
\end{align}
Using similar inequalities, \eqref{eq:second derivative of cost} can be written as
\begin{align}
\label{eq:second derivative of cost inequality}
\frac{d^2C_m}{dq^2}&\geq \frac{(2\mu-{\mu}^2)\lambda}{2{\mu}^3} \biggl( 2\int \limits_y \left( \mu_x + \rho_1(y-\mu_{y_1}) \right) Q(y-\mu_{y_1})f_{Y_1}(y) \, dy \notag \\
&+\frac{3q\lambda}{\mu} \int \limits_y \left( \mu_x + \rho_1(y-\mu_{y_1}) \right) Q(y-\mu_{y_1})f_{Y_1}(y) \, dy \notag \\
&+2\int \limits_y \left( \mu_x + \rho_2(y-\mu_{y_2}) \right) Q(y-\mu_{y_2})f_{Y_2}(y) \, dy \notag \\
&+\frac{3(1-q)\lambda}{\mu} \int \limits_y \left( \mu_x + \rho_2(y-\mu_{y_2}) \right) Q(y-\mu_{y_2})f_{Y_2}(y) \, dy \biggr)
\end{align}

$f_{Y_i}(y)$ is symmetric about $\mu_{y_i}$ with equal values on both the sides. $Q(y-\mu_{y_i})$ is a decreasing function for all y. As a result, \eqref{eq:first term inequality} is positive if $(\mu_x -\rho_i \mu_{y_i})$ takes sufficiently large value. A sufficient condition for this to happen is when $(\mu_x - \mu_{y_i})$ takes sufficiently large value for any $\mu_{y_i}$. For this particular example, where $\sigma_x=\sigma_{y_i}=1$, we find using numerical computation that a threshold for sufficiently large is given by $(\mu_x - \mu_{y_i}) > \frac{2}{3}$.
\end{IEEEproof}

\section{Proof for Proposition~\ref{prop3}}
\label{app:prop3}
We write
\begin{equation*}
\mathbb{E}[ XD \mid X>x^{\ast}]=\frac{1}{\mathbb{P}( X>x^{*})} \mathbb{E}[ XD \mathbf{1}_{\{X>x^{*}\}}]\mbox{.}
\end{equation*}
For a fixed $x^*$, $\mathbb{P}( X>x^{\ast})$ is a constant. The remaining term can be expanded as follows:
\begin{align}
\label{eq:Delay for high priority tasks}
\mathbb{E}[ XD \mathbf{1}_{\lbrace X>x^{\ast}\rbrace}]&=\mathbb{E}[ \mathbb{E}[ XD \mathbf{1}_{\lbrace X>x^{\ast} \rbrace} \mid X ] ] \notag \\
&=\int\limits_{x=x^{\ast}}^{\infty}x \mathbb{E}[ D \mid X=x] f_{X}(x) \, dx\mbox{.}
\end{align}

Recall that a task with priority $x$ is routed to the first agent with probability $p(x,\rho_1,\rho_2)$ and the second with probability $1-p(x,\rho_1,\rho_2)$. Therefore,
\begin{align}
\label{eq:Delay for a given task}
\mathbb{E}[ D \mid X=x ]&=p(x,\rho_1,\rho_2) \mathbb{E}[ D \mid X=x,1 ] \notag \\
&+( 1-p(x,\rho_1,\rho_2)) \mathbb{E}[ D \mid X=x,2]\mbox{,}
\end{align}
where $\mathbb{E}[ D \mid X=x,i]$ is the average delay of a task with priority $x$ given it is routed to the $i$th queue. For brevity, denote $\mathbb{E}[ D \mid X=x,i]$ as $E_{D,i}(x)$. Substituting \eqref{eq:Delay for a given task} in \eqref{eq:Delay for high priority tasks} gives
\begin{align}
\label{eq:Expression for delay for high priority tasks}
&\mathbb{E}[ XD \mathbf{1}_{\lbrace X>x^{\ast}\rbrace}] \\ \notag &=\int\limits_{x=x^{\ast}}^{\infty} x\left( p(x) \left( E_{D,1} (x)-E_{D,2} (x) \right) + E_{D,2} (x) \right) f_{X}(x)\mbox{.}
\end{align}

The minimization problem now becomes
\begin{equation}
\begin{aligned}
\label{eq:optimization for high priority tasks appendix}
& \min_{p(x)}
& & \int\limits_{x=x^{\ast}}^{\infty} x\left( p(x) \left( E_{D,1} (x)-E_{D,2} (x) \right) + E_{D,2} (x) \right) f_{X}(x)\\
& \mbox{s. t. }
& & 0 \leq p(x, \rho_1, \rho_2) \leq 1 \ \  \mbox{ for all } x \in \mathbb{R}\\
&
& & \mathbb{E}[ p(X, \rho_1, \rho_2)]=p^*\mbox{.}
\end{aligned}
\end{equation}
We first show that $E_{D,1}(x)$ depends on $p(X, \rho_1, \rho_2)$ only through $\mathbb{E}[p(X, \rho_1, \rho_2)]$. To do so, we expand $E_{D,1} \left( x \right)$ as follows:
\begin{align}
\label{eq:Average delay for a task in queue 1}
E_{D,1} \left( x \right)=& \mathbb{E}\left[ D \mid X=x,R=1 \right] \notag \\
=& \mathbb{E} \left[ \mathbb{E}\left[ D \mid X=x,R=1,Y_1 \right] \mid X=x,R=1 \right] \notag\\
=& \int\limits_{y=-\infty}^{\infty} \mathbb{E}\left[ D \mid X=x,Y_1=y,R=1 \right] f_{Y_1 \mid X,U}\left( y \mid x,1 \right)dx \notag\\
\stackrel{(a)}{=}& \int\limits_{y=-\infty}^{\infty} \mathbb{E}\left[ D \mid Y_1=y,R=1 \right] f_{Y_1 \mid X,R}\left( y \mid x,1 \right) \,dx \notag \\
=& \int\limits_{y} \left( \frac{2\mu - 2 \mu \lambda_{1,y} + \lambda_{1,y}^2}{2{\left( \mu - \lambda_{1,y} \right)}^2} \right) f_{Y_1  \mid X, R}(y \mid x, 1),
\end{align}
where $\{R=1\}$ denotes the first agent, $\lambda_{1,y}$ is the arrival rate of tasks to the first agent with priorities higher than $y$, and $(a)$ follows from the fact that given the $Y$ priority of a task, its average delay is independent of $X$. $\mathbb{E}\left[ D | X=x,Y_1=y,R=1 \right]$ is the average sojourn time for a task routed to the first queue.

Consider the conditional density function $f_{Y_1 | X, R}(y | x, 1)$. This can be written as
\[ 
f_{Y_1  \mid X, R}(y | x, 1)=\frac{f_{Y_1, X | R}(y, x | 1)}{f_{X | R}(x | 1)}\mbox{.} 
\]
The conditional joint distribution of $X$ and $Y_1$ given $R$ in the numerator is the same as the unconditional joint distribution. This is because $R=1$ is the unconditional routing probability and hence has the same probability for each task. Similarly the conditional distribution of $X$ given $R$ is the same as the unconditional distribution of $X$. Therefore $f_{Y_1  | X, R}(y | x, 1)$ is independent of $p(X, \rho_1, \rho_2)$. Also, as shown in Appendix~\ref{Proof of lemma 1}, $\lambda_{1,y}$ depends on $p(X, \rho_1, \rho_2)$ only through $\mathbb{E}[p(X, \rho_1, \rho_2)]$. Further, note that no other variable in \eqref{eq:Average delay for a task in queue 1} depends on $p(X, \rho_1, \rho_2)$. This implies that $E_{D,1}(x)$ depends on $p(X, \rho_1, \rho_2)$ only through $\mathbb{E}[p(X, \rho_1, \rho_2)]$.

Finally, note that minimizing \eqref{eq:optimization for high priority tasks appendix} is the same as minimizing its integrand at each $x \in ( x^*, \infty)$. Since $E_{D,1}(x)$ depends on $p(X, \rho_1, \rho_2)$ only through $\mathbb{E}[p(X, \rho_1, \rho_2)]$, the minima depends on the value of the coefficient of $p(x)$. Hence, the result follows.

\section{Cost function for Gaussian copula}
\label{app:copula}
We model $Y_1$ and $Y_2$ as the marginals of a Gaussian copula:
\begin{align}
\label{eq:copula again}
C_{\rho_G}(y_1,y_2)=\int \limits_{-\infty}^{{\Phi}^{-1}(y_1)} \int \limits_{-\infty}^{{\Phi}^{-1}(y_2)} \frac{1}{2\pi\sqrt{1-\rho_G^2}} \exp \left( -\frac{s^2 -2\rho_G st + t^2}{2(1-\rho_G^2)} \right) ds \, dt,
\end{align}
where ${\Phi}^{-1}(\cdot)$ is the quasi-inverse of standard normal distribution, and $\rho_G$ is the correlation in the bivariate normal distribution. Due to symmetry, the distribution of $Z_1$ and $Z_2$ will be the same.

Let us find the distribution of $Z_1$. We write
\begin{align}
\mathbb{P}(Z_1 \leq z)= & \mathbb{P}(Y_1 \leq z \mid Y_1 > Y_2) \notag \\
= & \frac{\mathbb{P}(Y_1 \leq z, Y_1 > Y_2)}{\mathbb{P}(Y_1 > Y_2)}.
\end{align}
We consider the case when $\rho_G=-0.8$. Using \eqref{eq:copula again},
\begin{align}
\label{eq:P(Y_1 < z,Y_1 > Y_2)}
&\mathbb{P}(Y_1 \leq z, Y_1 > Y_2) \notag \\
&= \int \limits_{y=-\infty}^{{\Phi}^{-1}(z)} \int \limits_{x=-\infty}^{y} \frac{1}{2\pi\sqrt{1-{\rho}^2}} \exp \left( -\frac{x^2 -2\rho xy + y^2}{2(1-{\rho}^2)} \right) dx \, dy \notag \\
&= \int \limits_{y=-\infty}^{{\Phi}^{-1}(z)} \frac{\left( 1+\erf \left( \frac{3}{\sqrt{2}}y \right) \right) \exp\left(-\frac{y^2}{2} \right)}{2\sqrt{2\pi}} dy.
\end{align}
Substituting $z=1$ in \eqref{eq:P(Y_1 < z,Y_1 > Y_2)} gives
\begin{equation*}
\mathbb{P}(Y_1 > Y_2)=\frac{1}{2}.
\end{equation*}
Therefore,
\begin{equation}
\label{eq:P(Z_1 < z)}
\mathbb{P}(Z_1 \leq z)=\int \limits_{y=-\infty}^{{\Phi}^{-1}(z)} \frac{\left( 1+\erf \left( \frac{3}{\sqrt{2}}y \right) \right) \exp\left(-\frac{y^2}{2} \right)}{\sqrt{2\pi}} dy.
\end{equation}
To find the density function, we differentiate the above equation. Letting $z=\Phi(\tau)$ and using Leibniz rule in \eqref{eq:P(Z_1 < z)} gives
\[ f_{Z_1}(z)=\frac{\frac{d{\Phi}^{-1}(z)}{dz} \left( 1+\erf \left( \frac{3}{\sqrt{2}}{\Phi}^{-1}(z) \right) \right) \exp\left(-\frac{{\left( {\Phi}^{-1}(z) \right)}^2}{2} \right)}{\sqrt{2\pi}}, \]
where
\begin{align}
\frac{d{\Phi}^{-1}(z)}{dz} =& \frac{1}{dz/d\tau} \notag \\
=& \sqrt{2\pi} \exp \left( \frac{{\left( {\Phi}^{-1}(z) \right)}^2}{2} \right).
\end{align}
As a result,
\[ f_{Z_1}(z)=\left( 1+\erf \left( \frac{3}{\sqrt{2}}{\Phi}^{-1}(z) \right) \right). \]
Similarly, we obtain the density function for $\rho_G=-0.4$ and $\rho_G=0$.
Using these density functions in \eqref{eq:conditional service time is geometric} gives
\begin{align*}
\mu_i&=0.68, \quad \text{var}(S_i)=2.48, \quad \rho_G=-0.8, \\
\mu_i&=0.59, \quad \text{var}(S_i)=2.73, \quad \rho_G=-0.4, \\
\mu_i&=0.53, \quad \text{var}(S_i)=2.99, \quad \rho_G=0.
\end{align*}
The cost can be easily computed using these values.

\section{Proof for Proposition~\ref{prop4}}
\label{app:prop4}
Consider the principal having $n$ agents to allocate tasks to. Let $Y_1, \ldots, Y_n$ be the random variables that denote the interests of the $n$ agents. We assume that the principal knows the agents' interest-realizations for each task. Let $\{ R=1 \}$ be the event that a task is routed to the first queue.
Since the agents work at variable rates, let $\mu(Y_k)$ be the conditional service rate for the $k$th agent. As earlier, we consider $\mu(Y_k)$ to be a concave function
and to be a common mapping across all agents $k$.

In the two-agent case, choosing agents with completely negatively correlated interests was advantageous, since it ensured maximum diversity in the agents' interests. In order to have complete diversity in the interest sets with $n$ agents, we create the following relation structure: the random variable $Y_k$ is distributed uniformly in $[ \frac{k-1}{n},\frac{k}{n}]$. Though highly simplified, this model is a good representation of the interests when the number of agents is large and the agents have fairly diverse interests. 

The routing policy is trivial: route a task to the agent who is interested in it. Let $Z_k$ be the random variable denoting the interests of the tasks in the queue of the $k$th agent. Then, the distribution for the $\{Z_k\}$ is the same as the $\{Y_k\}$:
\[ Z_k \sim \mathcal{U}\left( \frac{k-1}{n},\frac{k}{n} \right), \quad k=1, \ldots , n. \]
Let $\mathcal{M}_n$ denote the set of service rates for each agent when the total number of agents is $n$. Then,
\[ \mathcal{M}_n = \left( \mu\left(\frac{k-1}{n}\right),\mu\left( \frac{k}{n} \right) \right). \]
Since the agents have the same set of service rates,
\[ \mathcal{M}_n = \left( \mu\left(\frac{n-1}{n}\right),\mu\left( 1 \right) \right). \]
Note that $\mathcal{M}_{\infty}$, the set of service rates for each agent when the number of agents is infinite, is a singleton set given by $\mu(1)$. Denote by ${\widehat{\mathcal{M}}}_n$ the unconditional service rate for a pool of $n$ agents. Then,
\begin{equation}
\label{eq:unconditional service rate for n agents}
{\widehat{\mathcal{M}}}_{\infty} = \lim_{n \rightarrow \infty} {\widehat{\mathcal{M}}}_n=\mu(y)\bigg|_{y=1}.
\end{equation}

Let $C_{n}$ be the cost function for this case when there are $n$ agents. For brevity, denote $p_i=\mathbb{P}(R=i)$. $C_{n}$ expands as
\begin{align}
\label{eq:C_n}
C_{n}=\sum \limits_{i=1}^n \mathbb{E} [ XD | R=i]p_i.
\end{align}
By symmetry, it is clear
\[ p_i=\frac{1}{n}\mbox{, } i=1, \ldots, n\mbox{.} \]
As a result, \eqref{eq:C_n} becomes
\[ C_{n}=\frac{1}{n}\sum\limits_{i=1}^n \mathbb{E}[ XD | R=i ]\mbox{.} \]
From \eqref{eq:cost for one-agent deceitful agent one}, $\mathbb{E}[XD |R=i]$ can be expressed as
\begin{align}
\label{eq:cost for multi-agent deceitful agent i}
\mathbb{E}[XD \mid U=i]=\int \limits_x x f_X(x) dx \left(\frac{\left( 2\mu_i -{\lambda}_{i,x}\right)}{2{\left( \mu_i - {\lambda}_{i,x} \right)}^{2}} 
+ \frac{{\lambda}_{i,x} \mu_i^2 \mbox{var} \left( S_i \right)}{2{\left( \mu_i - {\lambda}_{i,x} \right)}^{2}} - \frac{{\lambda}_{i,x}}{2 \left( \mu_i - {\lambda}_{i,x} \right)}\right),
\end{align}
where $\mu_i$ is the unconditional service rate for the $i$th task and $\lambda_{i,x}$ is the arrival rate of tasks to the $i$th queue with priorities higher than $X=x$. Let $\lambda_x$ denote the arrival rate of tasks to the principal with priorities higher than $x$. Then,
\begin{equation}
\lambda_{i,x}=\lambda_x p_i =\frac{\lambda_x}{n}\mbox{.}
\end{equation}
Also, since $\mu_i$ is the unconditional service rate for the $i$th agent, it is the same for each agent, and is equal to ${\widehat{\mathcal{M}}}_n$. As a result,
\[ \mathbb{E}[XD | R=1]=\dots=\mathbb{E}[XD |R=n]\mbox{.} \]
Therefore, $C_{n}$ becomes
\[ C_{n}=\mathbb{E}[XD |R=1]\mbox{,} \]
where
\begin{align}
\label{eq:cost for multi-agent deceitful agent any}
\mathbb{E}[XD |R=1]=\int \limits_x x f_X(x) dx \left(\frac{\left( 2{\widehat{\mathcal{M}}}_n -\frac{\lambda_x}{n}\right)}{2{\left( {\widehat{\mathcal{M}}}_n - \frac{\lambda_x}{n} \right)}^{2}} 
+ \frac{\frac{\lambda_x}{n} {{\widehat{\mathcal{M}}}_n}^2 \mbox{var} \left( S_i \right)}{2{\left( {\widehat{\mathcal{M}}}_n - \frac{\lambda_x}{n} \right)}^{2}} - \frac{\frac{\lambda_x}{n}}{2 \left( {\widehat{\mathcal{M}}}_n - \frac{\lambda_x}{n} \right)}\right).
\end{align}
Taking the limit with the number of agents growing to infinity and using \eqref{eq:unconditional service rate for n agents} and \eqref{eq:cost for multi-agent deceitful agent any} gives
\begin{equation}
\label{eq:cost for infinite agents variable again}
C_{\infty}=\lim_{n \rightarrow \infty}C_{n}=\frac{1}{2\mu(y)}\bigg|_{y=1}.
\end{equation}

\bibliographystyle{IEEEtran}
\bibliography{conf_abrv,abrv,aseemref}

\end{document}